\documentclass[11pt,a4paper]{article}

\usepackage[utf8]{inputenc}
\usepackage[T1]{fontenc}
\usepackage{amsmath,amssymb,amsfonts}
\usepackage{graphicx}
\usepackage{hyperref}
\usepackage[margin=1in]{geometry}
\usepackage{amsthm}
\usepackage{algorithm}
\usepackage{algpseudocode}
\usepackage{natbib}
\usepackage[flushleft]{threeparttable}

\newenvironment{tablenotes}[1][]{\par\vspace{0.5em}\footnotesize\begin{trivlist}\item[]}{\end{trivlist}\normalsize}
\usepackage{booktabs}
\usepackage{multirow}
\usepackage{tabularx}
\usepackage{bbm}

\theoremstyle{plain}

\numberwithin{equation}{section}

\usepackage{array}
\newcolumntype{P}[1]{>{\centering\arraybackslash}p{#1}}
\begin{document}

\title{Deep $g$-Pricing for CSI 300 Index Options with Volatility Trajectories and Market Sentiment}

\author{
Yilun Zhang$^{1,2}$ \and Zheng Tang$^{3}$ \and Hexiang Sun$^{3}$ \and Yufeng Shi$^{1,3,4,*}$
}

\date{
\small $^{1}$Institute for Financial Studies, Shandong University, Jinan 250100, China \\
\small $^{2}$Research Center for Mathematics and Interdisciplinary Sciences, Shandong University, Qingdao 266237, China \\
\small $^{3}$School of Mathematics, Shandong University, Jinan 250100, China \\
\small $^{4}$State Key Laboratory of Cryptography and Digital Economy Security, Shandong University, Jinan 250100, China \\[0.5em]
\small $^{*}$Corresponding author: yfshi@sdu.edu.cn
}

\maketitle

\begin{abstract}
Option pricing in real markets faces fundamental challenges. The Black--Scholes--Merton (BSM) model assumes constant volatility and uses a linear generator $g(t,x,y,z)=-ry$, while lacking explicit behavioral factors, resulting in systematic departures from observed dynamics. This paper extends the BSM model by learning a nonlinear generator within a deep Forward--Backward Stochastic Differential Equation (FBSDE) framework. We propose a dual-network architecture where the value network $u_\theta$ learns option prices and the generator network $g_\phi$ characterizes the pricing mechanism, with the hedging strategy $Z_t=\sigma_t X_t \nabla_x u_\theta$ obtained via automatic differentiation. The framework adopts forward recursion from a learnable initial condition $Y_0=u_\theta(0,\cdot)$, naturally accommodating volatility trajectory and sentiment features. Empirical results on CSI 300 index options show that our method reduces Mean Absolute Error (MAE) by 32.2\% and Mean Absolute Percentage Error (MAPE) by 35.3\% compared with BSM. Interpretability analysis indicates that architectural improvements are effective across all option types, while the information advantage is asymmetric between calls and puts. Specifically, call option improvements are primarily driven by sentiment features, whereas put options show more balanced contributions from volatility trajectory and sentiment features. This finding aligns with economic intuition regarding option pricing mechanisms.
\end{abstract}

\vspace{1em}

\section{Introduction}
\label{sec:introduction}
Option pricing is one of the central problems in financial engineering. The classical Black--Scholes--Merton model \citep{CITE:Black_Scholes_1973,CITE:Merton_1973}, under the assumptions of no-arbitrage and market completeness, characterizes the dynamics of underlying asset prices via geometric Brownian motion and assumes constant volatility, thereby providing a closed-form solution for European option pricing. However, extensive empirical studies have documented systematic departures from these assumptions in real markets. Asset return distributions exhibit leptokurtosis and heavy tails, volatility displays pronounced clustering and leverage effects \citep{CITE:Cont_2001}, and implied volatility surfaces manifest smile or skew patterns \citep{CITE:Rubinstein_1994,CITE:Dumas_1998,CITE:Gatheral_2006}. These stylized facts indicate that the constant volatility assumption fails to effectively capture the true dynamics of financial markets. To address these limitations, the literature has developed various extended models. Local volatility models \citep{CITE:Dupire_1994,CITE:Derman_Kani_1994} specify volatility as a deterministic function of asset price and time, enabling approximate calibration to the current implied volatility surface; however, the absence of an independent stochastic volatility factor often renders them incapable of capturing the stochastic dynamics of the implied volatility surface. Stochastic volatility models \citep{CITE:Hull_White_1987,CITE:Heston_1993} introduce an additional stochastic process to describe volatility dynamics, effectively capturing volatility clustering and mean reversion, but at the cost of increased complexity in parameter calibration. All of the above approaches rely on pre-specified structural model assumptions and thus remain limited when confronting complex nonlinear pricing mechanisms.

Backward Stochastic Differential Equation (BSDE) theory provides a more general mathematical framework for option pricing, capable of characterizing broader pricing mechanisms within a unified framework. \citet{CITE:Pardoux_Peng_1990} established the existence and uniqueness theory for BSDE solutions, and \citet{CITE:Karoui_1997} systematically elaborated on the applications of BSDEs in finance, integrating derivative pricing and hedging into a unified framework. Within the BSDE framework, option value is jointly characterized by the terminal condition and the generator function, where the generator reflect the pricing mechanism of the market. The classical BSM model under the risk-neutral measure corresponds to a linear generator $g(t,y,z)=-ry$, where $r$ denotes the risk-free rate. In incomplete markets, however, the pricing operator often exhibits nonlinearity, corresponding to a nonlinear generator. Frictions such as model uncertainty \citep{CITE:Peng_1997}, transaction costs \citep{CITE:Leland_1985,CITE:Barles_Soner_1998}, and liquidity constraints \citep{CITE:Cetin_2004} can all be characterized through specific forms of nonlinear generators. The $g$-expectation theory proposed by \citet{CITE:Peng_g_expectation} further connects nonlinear expectations with BSDEs, providing a theoretical foundation for addressing complex market characteristics. Under this framework, option pricing can be transformed into a generator identification problem, that is, given the dynamics of the underlying asset, one can infer the generator that characterizes the pricing mechanism from market option price data, thereby converting the pricing problem into a machine learning problem. However, when the generator has complex nonlinear architectures, traditional numerical methods become computationally intractable, which creates an opportunity for the introduction of deep learning methods.

Deep learning has brought breakthroughs to the numerical solution of high-dimensional BSDEs. \citet{CITE:Han_Jentzen_2017,CITE:Han_2018} proposed the deep BSDE method, utilizing neural networks to approximate the solution of BSDEs and successfully solving problems with dimensions as high as 100 in numerical experiments. \citet{CITE:Hure_2020} further developed the deep backward scheme, improving numerical stability and convergence performance. \citet{CITE:Shi_2025} applied the deep BSDE method to option pricing, validating its effectiveness on real market data. However, most existing deep BSDE studies assume that the generator form is known or follows a rigid parametric architecture, which is overly idealized for real market pricing. Moreover, volatility is often assumed to be constant or modeled via oversimplified settings, failing to reflect the time-varying characteristics of the market. More importantly, existing methods generally share two common limitations. First, they lack an effective mechanism to infer unknown generators from market prices in real data scenarios. Second, a gap exists between model prediction and economic interpretation, making it difficult to attribute performance gains to specific financial mechanisms.

To address the above limitations, this paper considers incorporating richer market information to enhance the model's characterization capability. The proliferation of high-frequency data has made realized volatility (RV) an important tool for capturing market volatility dynamics \citep{CITE:Andersen_Bollerslev_1998,CITE:Andersen_2003,CITE:Barndorff_Nielsen_2002}, enabling timely reflection of changes in market conditions. Meanwhile, behavioral finance research has demonstrated that investor sentiment exerts a significant impact on asset prices \citep{CITE:Baker_Wurgler_2006}, and sentiment information extracted from media and online texts has been shown to be associated with stock market fluctuations \citep{CITE:Tetlock_2007,CITE:Antweiler_Frank_2004,CITE:Song_2025}. However, how to effectively integrate realized volatility and sentiment factors into the option pricing framework remains underexplored in the existing literature.

In summary, existing research still faces three limitations. First, volatility inputs are often constant or oversimplified. Second, behavioral finance factors such as investor sentiment are inadequately incorporated. Third, the internal mechanisms of deep models remain largely black-box, making it difficult to interpret the sources of performance improvement and the boundaries of applicability. To address these limitations, this paper proposes a deep $g$-pricing framework that integrates volatility trajectory features and market sentiment. The framework constructs a dual-network architecture comprising a value network and a generator network within the deep FBSDE architecture, explicitly computes the hedging ratio via automatic differentiation, and introduces a gating mechanism to achieve adaptive fusion of volatility trajectory and sentiment features.

The main contributions of this paper are threefold. First, at the architectural level, we propose a dual-network architecture comprising a value network and a generator network, explicitly compute the hedging ratio via automatic differentiation, and employ a pre-training and fine-tuning strategy to enhance training stability and generalization capability. Second, at the information fusion level, we incorporate predicted volatility trajectories and sentiment features into the FBSDE recursion, achieving adaptive information utilization through differentiated gating mechanisms. Third, at the interpretability level, we construct an analytical closed-loop spanning from error advantage decomposition to information contribution decomposition and finally to sample-level testing, revealing the asymmetry in information dependence between call and put options.

The remainder of this paper is organized as follows. Section~\ref{sec:methodology} introduces the research methodology and modeling framework. Section~\ref{sec:experiments} presents the empirical design and results. Section~\ref{sec:xai} conducts the interpretability analysis. Section~\ref{sec:conclusion} concludes the paper and discusses future directions.

\section{Methodology}
\label{sec:methodology}

\subsection{Realized Volatility Prediction}
\label{subsec:rv}

Realized volatility, proposed by \citet{CITE:Andersen_Bollerslev_1998}, is a volatility measure computed as the sum of squared high-frequency returns. Let $r_{t,i}$ denote the $i$-th high-frequency log return on day $t$, and let $M$ denote the number of high-frequency observations on that day. The realized volatility is defined as
\begin{equation}
	RV_t = \sum_{i=1}^{M} r_{t,i}^2,
\end{equation}
Strictly speaking, $RV_t$ defined by the above equation represents realized variance. In the option pricing model of this paper, the annualized volatility input $\sigma_t$ is converted via $\sigma_t = \sqrt{RV_t} \times \sqrt{252}$, where 252 denotes the number of trading days per year. For simplicity, we henceforth use RV to refer to realized volatility throughout the paper.

Regarding the forecasting methodology, this paper builds upon our team's prior research \citep{CITE:Zhang_2025}, using RV computed from high-frequency data as the prediction target and employing deep learning models for modeling. Specifically, the model comprises two types of feature extraction modules. First, large language models are utilized to perform sentiment analysis on investor comments, extracting textual features that reflect market expectations. Second, market data such as prices and trading volumes are transformed into image representations, from which volatility intensity patterns are extracted via convolutional neural networks \citep{CITE:LeCun_1998_CNN}. Building upon this foundation, the TimeXer framework \citep{CITE:Wang_TimeXer_2024} integrates self-attention and cross-attention mechanisms \citep{CITE:Vaswani_2017_Attention} to simultaneously capture the temporal dependence of RV and external feature information, thereby enabling the prediction of future volatility. For detailed architecture and performance evaluation of the RV forecasting model, we refer readers to our team's prior work; this paper directly uses its output as the volatility trajectory input for option pricing. It should be emphasized that the volatility $\sigma_t$ used in the option pricing model of this paper represents the predicted future RV trajectory generated by the aforementioned forecasting model, rather than historical realized volatility. Compared with the traditional constant volatility assumption, this approach provides dynamic volatility trajectory inputs for option pricing.

To avoid look-ahead bias, this paper strictly adheres to the principle of information availability for temporal alignment. For each pricing date $t=0$, the RV forecasting model constructs features using only historical data available on or before that date, generating daily volatility trajectory forecasts over the future interval $[0,\tau]$, with no post-expiration information used in the prediction process.

\subsection{Sentiment Feature Construction}
\label{subsec:sentiment}

The sentiment features in this paper are constructed based on post titles from Guba (a Chinese stock forum). Compared with comments, post titles more directly reflect the subjective sentiment of the posters, involve higher composition costs, and exhibit more concise expression with stronger summarization and representativeness. Accordingly, this paper collected all post titles from 2016 to 2024 and employed a two-stage approach for sentiment classification. In the first stage, the BERT model \citep{CITE:Devlin_2019_BERT} was used for preliminary screening of titles to filter out noisy samples. In the second stage, multiple large language models, including DISC-FinLLM, Qwen, and DeepSeek, were employed for refined classification, with voting across model outputs to ensure accuracy. Ultimately, each title was classified into one of three categories: bullish, bearish, or neutral. Based on this classification, this paper computes daily statistics including the number of bullish titles $N_{bull,t}$, the number of bearish titles $N_{bear,t}$, total number of posts, total page views, and total number of comments, from which multidimensional sentiment features are constructed to characterize the directionality and uncertainty of investor expectations.

For options with different moneyness levels, this paper constructs differentiated sentiment feature sets. For at-the-money options, short-term sentiment indicators are adopted to capture the sensitivity in the price critical region. For out-of-the-money options, medium-term sentiment indicators and extreme event flags are employed to characterize tail risk sensitivity. Table~\ref{tab:sentiment_features} provides a detailed list of the sentiment feature indicators used for each option category, along with their formulas and economic interpretations.

\begin{table}[htbp]
\caption{Feature Construction Based on Sentiment and Activity}\label{tab:sentiment_features}
\scriptsize
\renewcommand{\arraystretch}{1.3}
\begin{tabularx}{\textwidth}{@{} p{1.2cm} p{1.8cm} P{5.5cm} X @{}}
\toprule
\textbf{Option} & \textbf{Features} & \textbf{Formula} & \textbf{Explanation} \\
\midrule
ATM & NetSent
& $\mathrm{NetSent}_t=\frac{N_{\mathrm{bull},t}-N_{\mathrm{bear},t}}{\mathrm{posts\_count}_t+\varepsilon}$
& Overall bullish--bearish tilt on day $t$ \\

ATM & Surprise$_3$
& $\mathrm{NetSent}_t-\mathrm{EWM}_{h=3}(\mathrm{NetSent})_t$
& Short-term deviation from recent sentiment baseline \\

ATM & Disp$_3$
& $\mathrm{EWMstd}_{h=3}(\mathrm{NetSent})_t$
& Short-term sentiment volatility, reflecting disagreement \\

ATM & zActivity
& $z\mathrm{Act}_t=\frac{\mathrm{Activity}_t-\mathrm{EWM}_{h=10}(\mathrm{Activity})_t}{\mathrm{EWMstd}_{h=10}(\mathrm{Activity})_t+\varepsilon}$
& Abnormal discussion intensity, often aligned with event windows \\

ATM & Entropy
& $\mathrm{Entropy}_t=-\!\big[p^{(\mathrm{bull})}_t\ln p^{(\mathrm{bull})}_t+(1-p^{(\mathrm{bull})}_t)\ln(1-p^{(\mathrm{bull})}_t)\big]$
& Balance of opinions, maximized when bulls and bears are evenly split \\

\midrule
OTM & Surprise$_5$
& $\mathrm{NetSent}_t-\mathrm{EWM}_{h=5}(\mathrm{NetSent})_t$
& Medium-term deviation from sentiment baseline \\

OTM & Disp$_5$
& $\mathrm{EWMstd}_{h=5}(\mathrm{NetSent})_t$
& Medium-term sentiment volatility, indicating persistent disagreement \\

OTM & Extreme\_flag
& $\mathrm{Extreme\_flag}_t=\mathbbm{1}\!\big(|z^{(30)}_t|>2.0\big)$
& Rare extreme sentiment events relative to a long-horizon baseline \\

OTM & VolShock\_flag
& $\mathrm{VolShock\_flag}_t=\mathbbm{1}\!\big(z\mathrm{Act}_t>1.5\big)$
& Bursts of abnormal activity, indicating potential information shocks  \\

OTM & EWM\_NetSent
& $\mathrm{EWM}_{h=5}(\mathrm{NetSent})_t$
& Short-term sentiment momentum, capturing potential directional shifts \\
\bottomrule
\end{tabularx}

\begin{tablenotes}[flushleft]
\scriptsize
\item \textbf{Notes:}
$\mathrm{EWM}_{h=\cdot}$ and $\mathrm{EWMstd}_{h=\cdot}$ denote exponentially weighted mean and standard deviation with half-life $h$;
$\mathrm{Activity}_t = \ln(1+\text{posts\_count}_t) + \ln(1+\text{views\_sum}_t) + \ln(1+\text{comments\_sum}_t);$
$\varepsilon$ is a small stabilizer;
$\,p^{(\mathrm{bull})}_t = \frac{N_{\mathrm{bull},t}}{\mathrm{posts\_count}_t+\varepsilon};$
$\,z^{(30)}_t = \frac{\mathrm{NetSent}_t-\mathrm{EWM}_{h=30}(\mathrm{NetSent})_t}{\mathrm{EWMstd}_{h=30}(\mathrm{NetSent})_t+\varepsilon}.$
\end{tablenotes}
\end{table}

\subsection{Deep Learning-Based $g$-Pricing Method}
\label{subsec:fbsde}

\subsubsection{FBSDE Neural Network Formulation}
\label{subsubsec:fbsde_nn_formulation}
Consider an FBSDE system on the probability space $(\Omega, \mathcal{F}, \mathbb{Q})$, where $\mathbb{Q}$ denotes the risk-neutral pricing measure and $\{\mathcal{F}_t\}_{t \in [0,\tau]}$ is the natural filtration generated by the $\mathbb{Q}$-Brownian motion $\{W_t\}_{t \in [0,\tau]}$. The underlying asset price $X_t$ follows the dynamics under $\mathbb{Q}$:
\begin{equation}
\begin{cases}
	dX_t = rX_t\, dt + \sigma_t X_t\, dW_t, \quad X_0 = x_0, \\
	dY_t = -g(t, X_t, Y_t, Z_t, \mathbf{e}_t)\, dt + Z_t\, dW_t, \quad Y_\tau = \Phi(X_\tau),
\end{cases}
\end{equation}
where $r$ denotes the risk-free rate, $\sigma_t$ represents the annualized volatility trajectory feature generated by the RV forecasting model, and $\mathbf{e}_t \in \mathbb{R}^5$ is the market sentiment feature vector. Since sentiment at time $t>0$ is unavailable on the pricing date, we set $\mathbf{e}_t \equiv \mathbf{e}_0$ for $t \in [0,\tau]$ in the empirical implementation. On each pricing date, $\{\sigma_t\}_{t\in[0,\tau]}$ and $\mathbf{e}_0$ are treated as $\mathcal{F}_0$-measurable exogenous inputs generated from information available on the pricing date, and the FBSDE system is solved conditional on these inputs. This paper adopts the setting $r=0$ to maintain consistency with the Tengbinn baseline. For European options, the terminal condition is $\Phi(X_\tau) = (X_\tau - K)^+$ for call options or $\Phi(X_\tau) = (K - X_\tau)^+$ for put options, where $K$ denotes the strike price.

The core of this method lies in using two neural networks to approximate the value function and the generator, respectively. The value network $u_\theta(t, X, K, \tau, \sigma, r, \mathbf{e}): \mathbb{R}^{11} \to \mathbb{R}$ predicts the option value, while the generator network $g_\phi(t, X, Y, Z, K, \tau, \sigma, r, \mathbf{e}): \mathbb{R}^{13} \to \mathbb{R}$ characterizes the pricing mechanism, where $\theta$ and $\phi$ denote the parameters of the two networks, respectively. According to the nonlinear Feynman--Kac formula \citep{CITE:Feynman_1948,CITE:Kac_1949}, the hedging strategy can be computed via automatic differentiation:
\begin{equation}
Z_t = \sigma_t X_t \frac{\partial u_\theta}{\partial X}(t, X_t, \cdot).
\end{equation}

The generator network $g_\phi$ in this paper takes contract parameters $(K, \tau)$ and the volatility trajectory $\sigma_t$ as inputs, which differs from the standard BSDE theory where the generator depends only on $(t, X, Y, Z)$. This design can be understood from two perspectives.

First, from the contract-conditioned generator perspective, $g_\phi(t, X_t, Y_t, Z_t; K, \tau, \sigma_t, r, \mathbf{e}_t)$ can be interpreted as an effective generator conditional on contract specifications $(K,\tau)$, characterizing the local pricing mechanism across different moneyness levels and maturities, encompassing contract-specific factors such as liquidity premium, demand pressure, and tail risk.

Second, from the extended state space perspective, $(K, \tau, \sigma_t, r, \mathbf{e}_t)$ can be regarded as components of an extended state. Defining $\tilde{X}_t = (X_t, K, \tau, \sigma_t, r, \mathbf{e}_t)$, the generator retains the standard form $g_\phi(t, \tilde{X}_t, Y_t, Z_t)$. This is mathematically equivalent to incorporating contract parameters and market state into a unified state vector.

This paper adopts the contract-conditioned perspective for modeling. This implies that the learned generator represents an effective pricing mechanism tailored to specific contract types, rather than a universal market generator. This modeling choice aligns more closely with empirical pricing requirements, though the model's transferability is limited to the range of contracts covered by the training data. From a theoretical standpoint, the primary objective of this paper is empirical pricing accuracy and mechanism insight. The learned $g_\phi$ serves the purpose of pricing calibration for specific contract types, and its scope of applicability differs from that of the universal generator in standard $g$-expectation theory.

\subsubsection{Time Discretization and Numerical Scheme}
\label{subsubsec:time_discretization_scheme}

For numerical implementation, the time interval $[0,\tau]$ is uniformly partitioned into $N$ subintervals with time step $\Delta t = \tau/N$ and discrete time points $t_i = i\Delta t$ for $i = 0,1,\ldots,N$. Both $\tau$ and $\Delta t$ are measured in years, consistent with the dimension of the annualized volatility $\sigma_t$. For the forward SDE, this paper employs the analytical solution form of geometric Brownian motion for discretization to preserve price positivity. For the backward SDE, the Euler scheme is adopted for recursion \citep{CITE:Kloeden_Platen_1992}.

This paper employs a forward recursion scheme starting from $Y_0$, rather than the backward recursion used in traditional BSDE solvers. Specifically, the value network $u_\theta$ provides the initial value $Y_0 = u_\theta(0, X_0, \cdot)$, which then evolves forward in time, with consistency to the theoretical BSDE solution ensured through terminal condition loss and path consistency loss.

The discretization scheme for the forward process is
\begin{equation}
X_{t_{i+1}} = X_{t_i} \exp\left[\left(r - \frac{\sigma_{t_i}^2}{2}\right)\Delta t + \sigma_{t_i}\sqrt{\Delta t}\,\xi_{i+1}\right],
\end{equation}
The backward process adopts the forward recursion scheme
\begin{equation}
Y_{t_{i+1}} = Y_{t_i} - g_\phi(t_i, X_{t_i}, Y_{t_i}, Z_{t_i}, \cdot)\Delta t + Z_{t_i}\Delta W_{i+1},
\end{equation}
where $\Delta W_{i+1} = \sqrt{\Delta t}\,\xi_{i+1}$ and $\xi_{i+1} \sim \mathcal{N}(0,1)$ are independent standard normal random variables. This discretization scheme is consistent with the sign convention of the continuous-time BSDE $dY_t = -g\,dt + Z_t\,dW_t$.

\subsubsection{Loss Functions}
\label{subsubsec:loss_functions}

To ensure that the neural networks correctly learn the solution of the FBSDE, this paper designs a loss function comprising three components. The terminal loss and path loss adopt the RMSE form to maintain gradient scales consistent with price dimensions. The initial value loss adopts a hybrid form of RMSE and MAPE, balancing both absolute and relative errors.

First, the initial value matching loss adopts a hybrid form of RMSE and MAPE
\begin{equation}
\mathcal{L}_{\text{price}} = \omega_r \cdot \text{RMSE}(\hat{Y}_0, Y_0^{\text{market}}) + \omega_m \cdot \text{MAPE}(\hat{Y}_0, Y_0^{\text{market}}),
\end{equation}
where $\hat{Y}_0 = u_0 = u_\theta(0, X_0, \cdot)$ denotes the model-predicted initial value from the value network, and $\omega_r$ and $\omega_m$ are hyperparameters for the hybrid weights. When computing MAPE, a lower bound truncation is applied to the denominator to avoid numerical instability for zero-price samples, specifically using the form $\max(|y|, \epsilon)$, where $\epsilon$ is a small positive number. The hybrid loss design enables the model to simultaneously focus on absolute and relative errors, achieving good fitting performance for both high-priced and low-priced options.

Second, the terminal condition loss is defined as
\begin{equation}
\mathcal{L}_{\text{terminal}} = \text{RMSE}\left(u_\theta(\tau, X_\tau, \cdot), \Phi(X_\tau)\right),
\end{equation}
where $\Phi(X_\tau)$ denotes the option payoff function at expiration. The terminal loss adopts the RMSE form to avoid numerical instability of MAPE when the payoff approaches zero.

Third, the path consistency loss is defined as
\begin{equation}
\mathcal{L}_{\text{path}} = \text{RMSE}\left(U_{\text{path}}^{[0:N-1]}, Y_{\text{path}}^{[0:N-1]}\right),
\end{equation}
where $U_{\text{path}} = \{u_\theta(t_i, X_{t_i}, \cdot)\}_{i=0}^{N-1}$ represents the direct evaluation of the value network at each time step, and $Y_{\text{path}} = \{Y_{t_i}\}_{i=0}^{N-1}$ is computed through the BSDE recursion formula. This loss term constrains the value function to satisfy the theoretical definition $Y_{t_i} = u(t_i, X_{t_i})$, requiring that the function learned by the value network remains consistent with the BSDE recursion path, which is a core requirement for the theoretical solution of FBSDEs.

The total loss function is a weighted combination of the above three terms
\begin{equation}
\mathcal{L} = \lambda_1 \mathcal{L}_{\text{price}} + \lambda_2 \mathcal{L}_{\text{terminal}} + \lambda_3 \mathcal{L}_{\text{path}},
\end{equation}
where $\lambda_1, \lambda_2, \lambda_3$ are hyperparameters controlling the relative importance of each loss term. In practice, expected values are estimated via Monte Carlo methods. From a financial perspective, $\mathcal{L}_{\text{price}}$ achieves cross-sectional calibration, enabling the model to fit current market quotes. $\mathcal{L}_{\text{terminal}}$ guarantees the no-arbitrage terminal condition, ensuring that the payoff at expiration is consistent with the contract definition. $\mathcal{L}_{\text{path}}$ enforces temporal consistency, requiring the value function to satisfy the BSDE dynamics throughout the entire pricing interval.

\subsubsection{Neural Network Architecture}
\label{subsubsec:nn_architecture}

This paper adopts an adaptive activation function network architecture \citep{CITE:Agostinelli_2014}, combining five activation functions with adaptively learned weights to enhance expressive power.

To fully extract semantic features from financial variables, the network first independently expands each input variable from 1 dimension to 50 dimensions
\begin{equation}
h_i = W_i x_i + b_i, \quad W_i \in \mathbb{R}^{50 \times 1}, \quad i \in \{t, X, K, \tau, \sigma, r\},
\end{equation}
where $t$ denotes the current time, $X$ the underlying asset price, $K$ the strike price, $\tau$ the time to maturity, $\sigma$ the volatility, and $r$ the risk-free rate. The variable $r$ is retained as an input channel in the network architecture to maintain generality. The expansion layer weights are initialized using Xavier initialization \citep{CITE:Glorot_Bengio_2010_Xavier} to control output scale, with a gain coefficient of 0.01 and biases initialized to 0. The six expanded feature vectors are concatenated to form a 300-dimensional base financial representation.

The sentiment gating mechanism employs a three-step design. The sentiment feature vector $\mathbf{e}_t \in \mathbb{R}^5$ contains five market sentiment indicators. For options with different moneyness levels, this paper adopts differentiated sentiment feature sets: at-the-money options use short-term sentiment indicators, while out-of-the-money options use medium-term sentiment indicators and extreme event flags. Detailed indicator definitions are provided in Table~\ref{tab:sentiment_features}.

First, the sentiment embedding layer expands the 5-dimensional sentiment features to 50 dimensions
\begin{equation}
h_{\text{sent}} = W_s \mathbf{e}_t + b_s, \quad W_s \in \mathbb{R}^{50 \times 5}.
\end{equation}

Second, global gating modulates the contribution of sentiment features through a learnable scalar $\gamma$
\begin{equation}
\tilde{h}_{\text{sent}} = \gamma \cdot h_{\text{sent}},
\end{equation}
where $\gamma$ is initialized with different priors according to option moneyness and optimized jointly with other network parameters during training.

Third, the gated sentiment features are concatenated with the financial features
\begin{equation}
h_{\text{in}} = [h_{\text{fin}}; \tilde{h}_{\text{sent}}],
\end{equation}
where $h_{\text{fin}} \in \mathbb{R}^{300}$ is the concatenation of expanded features from the six financial variables, resulting in a final input dimension of 350.

The hidden layers of the network employ adaptive activation functions, combining five base activation functions with L2-normalized weights \citep{CITE:Hendrycks_Gimpel_2016_GELU,CITE:Ramachandran_2017_Swish} to ensure scale invariance of activation intensity
\begin{equation}
\begin{split}
\text{AAF}(x; \mathbf{w}) = \frac{1}{\|\mathbf{w}\|_2}
\bigg( & a \sin(x) + b \tanh(x) + c \cdot \text{GELU}(x)  + d \cdot \text{SiLU}(x) + e \cdot \text{Softplus}(x) \bigg),
\end{split}
\end{equation}
where the weight vector $\mathbf{w} = [a, b, c, d, e]^\top$ is initialized with equal weights $[0.2, 0.2, 0.2, 0.2, 0.2]^\top$ and adaptively learned through gradient descent during training. The motivation for adopting adaptive activation functions lies in the significant curvature differences and nonlinear architectures exhibited by option pricing functions across different moneyness levels, such as local sensitivity near at-the-money regions and the characteristics of low prices with high relative errors in out-of-the-money regions. Such nonlinearities may amplify gradient instability during forward recursion training of FBSDEs. Adaptive activation functions provide an adaptive nonlinear basis for the network by learning combinations of multiple base activation functions, enabling the model to automatically select appropriate nonlinear shapes across different state regions, thereby enhancing expressive power and convergence stability with minimal structural overhead. Weight normalization further controls activation scale and reduces training oscillations.

\begin{figure}[htbp]
	\centering
	\includegraphics[width=0.95\textwidth]{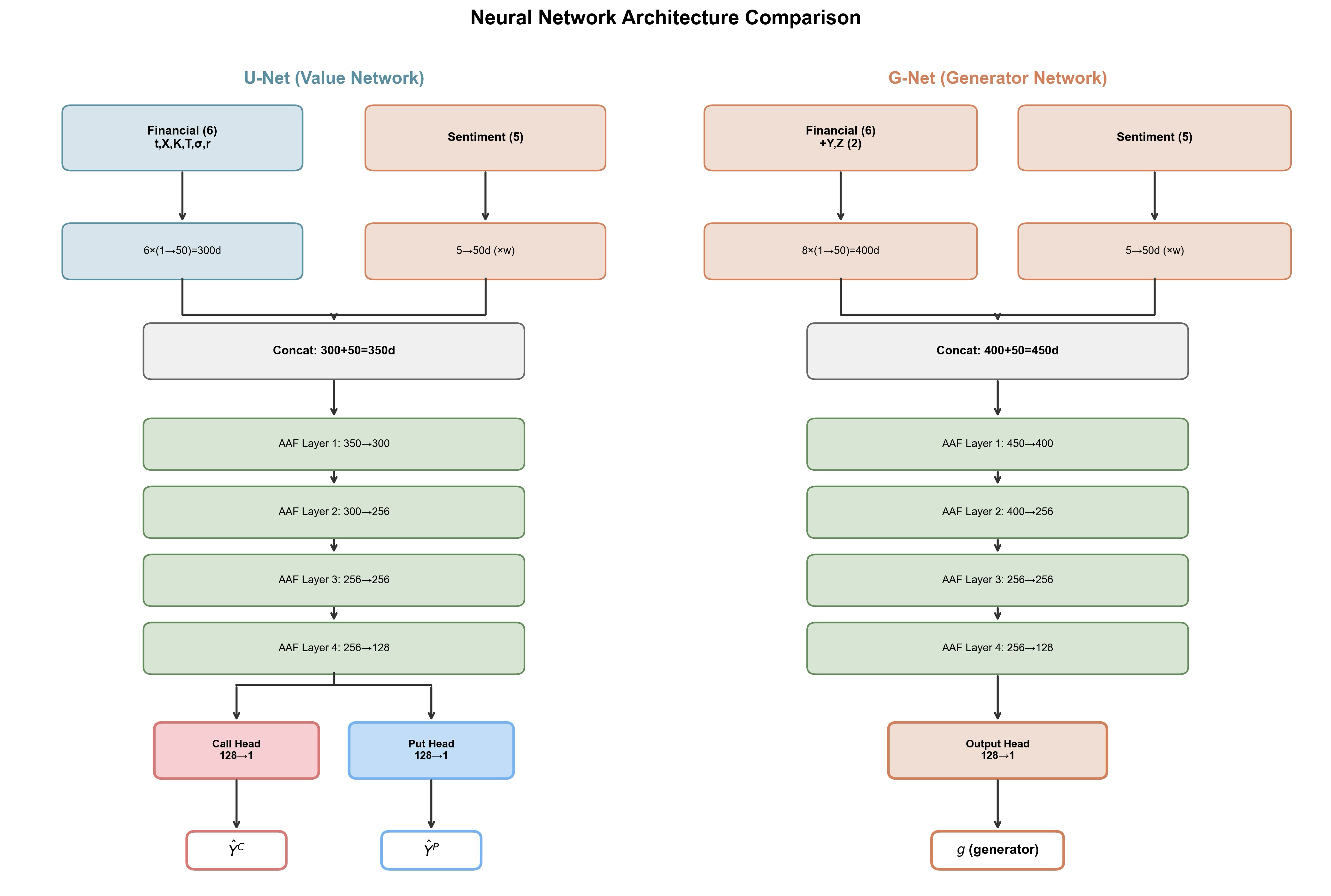}
	\caption{Architecture of the value network and generator network. The left panel shows the value network, illustrating the financial variable expansion layer, sentiment feature embedding layer with learnable gating parameter, shared backbone, and dual-head output architecture. The right panel shows the generator network, which has a similar architecture but additionally receives $Y_t$ and $Z_t$ as inputs.}
	\label{fig:network_arch}
\end{figure}

The overall network architecture is illustrated in Figure~\ref{fig:network_arch}. The value network adopts a shared backbone with dual-head output design. Each of the six financial variables is expanded from 1 dimension to 50 dimensions, while sentiment features are expanded to 50 dimensions through an embedding layer and modulated by a learnable gating parameter $w$, forming a 350-dimensional input after concatenation. The shared backbone consists of four adaptive activation function layers with dimensions of 350, 300, 256, 256, and 128 respectively, ultimately outputting predicted option values through separate call and put heads.

The generator network adopts a similar architecture but additionally receives the current option value $Y_t$ and hedging position $Z_t$ as inputs for computing the generator function in the FBSDE recursion. Its input comprises eight financial variables and five-dimensional sentiment features, which after expansion form a 450-dimensional input, passing through four adaptive activation function layers to output the generator value $g$.

The value network adopts a shared backbone with dual-head output architecture to distinguish the pricing characteristics of call and put options
\begin{equation}
\hat{Y}_0^{\text{call}} = \text{Head}_{\text{call}}(h), \quad \hat{Y}_0^{\text{put}} = \text{Head}_{\text{put}}(h),
\end{equation}
where $h$ denotes the output features from the shared backbone, and $\text{Head}_{\text{call}}$ and $\text{Head}_{\text{put}}$ are the independent output heads for call and put options, respectively. During inference, the result from the corresponding output head is selected based on the option type. This design allows the shared backbone to learn pricing architectures common to both call and put options, while the two output heads learn type-specific pricing mechanisms, achieving a balance between parameter sharing and task specialization.

\subsubsection{Training Strategy}
\label{subsubsec:training_strategy}

The training process adopts an end-to-end approach. The complete algorithm is presented in Algorithm~\ref{alg:deep_learning_g_pricing} and consists of the following main steps.

First, for each batch, $M$ independent sequences of Brownian motion increments $\{\Delta W_{t_i}\}$ are generated.

Second, the underlying asset price paths $\{X_{t_i}\}_{i=0}^N$ are computed through numerical recursion of the geometric Brownian motion SDE
\begin{equation}
X_{t_{i+1}} = X_{t_i} \exp\left((r-\frac{\sigma_{t_i}^2}{2})\Delta t + \sigma_{t_i}\Delta W_{t_{i+1}}\right).
\end{equation}

Third, the value network $u_\theta$ is used to directly evaluate the option value at each time step, generating the value network path $\{U_{t_i}\}_{i=0}^N$, where $U_{t_i} = u_\theta(t_i, X_{t_i}, \cdot)$.

Fourth, initializing $Y_0 = U_0$, the recursion path $\{Y_{t_i}\}_{i=0}^N$ is computed through forward BSDE recursion
\begin{equation}
Y_{t_{i+1}} = Y_{t_i} - g_\phi(t_i, X_{t_i}, Y_{t_i}, Z_{t_i}, \cdot)\Delta t + Z_{t_i}\Delta W_{t_{i+1}},
\end{equation}
where $Z_{t_i} = \sigma_{t_i} X_{t_i} \nabla_X u_\theta(t_i, X_{t_i}, \cdot)$ is computed via automatic differentiation.

Fifth, the three loss terms are calculated and network parameters are updated through backpropagation. The initial value matching loss $\mathcal{L}_{\text{price}}$ compares $U_0$ with market prices, the terminal condition loss $\mathcal{L}_{\text{terminal}}$ compares $U_\tau$ with the payoff at expiration, and the path consistency loss $\mathcal{L}_{\text{path}}$ constrains the value function to satisfy $U_{t_i} \approx Y_{t_i}$.

The optimizer is AdamW, the learning rate schedule adopts cosine annealing, and Dropout regularization is applied to improve generalization. Detailed hyperparameter configurations are provided in Table~\ref{tab:hyper}.

This method adopts a shared backbone with dual-head output architecture, and training is divided into two stages: pre-training and fine-tuning. In the pre-training stage, the shared backbone and both output heads are trained simultaneously, enabling the model to learn general pricing architectures. In the fine-tuning stage, the corresponding output head parameters and sentiment gating parameter $w$ are updated on each moneyness-specific subset, where $w$ for in-the-money options is fixed at 0 and not updated, and the shared backbone parameters are either frozen or fine-tuned with a smaller learning rate.

The above two-stage training strategy can be understood from the perspective of curriculum learning \citep{CITE:Bengio_2009_Curriculum}. The pre-training stage uses mixed option data to learn general patterns of option pricing, corresponding to the easy task. The fine-tuning stage learns differentiated pricing mechanisms for options with specific moneyness levels, corresponding to the difficult task. This training sequence from easy to difficult, from general to specialized, helps the model establish robust parameter initialization, thereby achieving faster convergence and improved generalization in the fine-tuning stage.

\begin{algorithm}[htbp]
	\caption{Deep Learning $g$-Pricing Algorithm}
	\label{alg:deep_learning_g_pricing}
	\begin{algorithmic}[1]
		\Require Underlying price $X_0$, time to maturity $\tau$, strike price $K$, risk-free rate $r$, known (call or put) option market price $Y_0^{\star}$, time discretization $\Delta t = \tau/N$, Brownian motion increment paths $\{\Delta W_{t_{i+1}}^j\}_{0 \leq i \leq N-1, 1 \leq j \leq M}$, volatility trajectory $\{\sigma_{t_i}\}$, sentiment features $\mathbf{e}_{t_i}$ (empirically $\mathbf{e}_{t_i} \equiv \mathbf{e}_0$), loss weights $\lambda_1, \lambda_2, \lambda_3$, hybrid loss weights $\omega_r, \omega_m$, and learning rate $\eta$
		\Ensure $u_0 = u_\theta(0, x_0, K, \tau, \sigma_0, r, \mathbf{e}_0)$

		\State $X_0^j \leftarrow x_0$ for $j = 1, 2, \ldots, M$
		\For{$i = 0, 1, \ldots, N-1$}
		\State $X_{t_{i+1}}^j \leftarrow X_{t_i}^j \exp\left( \left(r - \frac{1}{2} \sigma_{t_i}^2\right) \Delta t + \sigma_{t_i} \cdot \Delta W_{t_{i+1}}^j \right)$ \Comment{Forward recursion}
		\EndFor

		\State $u_0^j \leftarrow u_\theta(0, X_0^j, K, \tau, \sigma_0, r, \mathbf{e}_0)$ for $j = 1, 2, \ldots, M$ \Comment{Value function initial value (learnable)}
		\State $Y_0^j \leftarrow u_0^j$ for $j = 1, 2, \ldots, M$ \Comment{Initialize $Y_0$ for BSDE recursion}

		\For{$i = 0, 1, \ldots, N-1$}
		\State $u_{t_i}^j \leftarrow u_\theta(t_i, X_{t_i}^j, K, \tau, \sigma_{t_i}, r, \mathbf{e}_{t_i})$ \Comment{Value network direct evaluation}
		\State $Z_{t_i}^j \leftarrow \sigma_{t_i} X_{t_i}^j \nabla_X u_{t_i}^j$ \Comment{Hedging term: $\nabla_X u$ via automatic differentiation}
		\State $g_{t_i}^j \leftarrow g_\phi(t_i, X_{t_i}^j, Y_{t_i}^j, Z_{t_i}^j, K, \tau, \sigma_{t_i}, r, \mathbf{e}_{t_i})$ \Comment{Generator network}
		\State $Y_{t_{i+1}}^j \leftarrow Y_{t_i}^j - g_{t_i}^j \Delta t + Z_{t_i}^j \cdot \Delta W_{t_{i+1}}^j$ \Comment{BSDE recursion: $dY=-g\,dt+Z\,dW$}
		\EndFor
		\State $u_\tau^j \leftarrow u_\theta(\tau, X_\tau^j, K, \tau, \sigma_\tau, r, \mathbf{e}_\tau)$ for $j = 1, 2, \ldots, M$ \Comment{Terminal value function evaluation}

		\State $L_{\text{price}} \leftarrow \omega_r \cdot \text{RMSE}(\{u_0^j\}, Y_0^{\star}) + \omega_m \cdot \text{MAPE}(\{u_0^j\}, Y_0^{\star})$ \Comment{Hybrid loss}
		\State $L_{\text{terminal}} \leftarrow \text{RMSE}(\{u_\tau^j\}, \{\Phi(X_\tau^j)\})$
		\State $L_{\text{path}} \leftarrow \text{RMSE}(\{u_{t_i}^j\}_{i<N}, \{Y_{t_i}^j\}_{i<N})$
		\State $\mathcal{L} \leftarrow \lambda_1 L_{\text{price}} + \lambda_2 L_{\text{terminal}} + \lambda_3 L_{\text{path}}$

		\State $\theta \leftarrow \theta - \eta \nabla_{\theta} \mathcal{L}$ \Comment{Neural network parameters $\theta = \{\theta^{[u]}, \phi^{[g]}\}$ updated by gradient descent}
	\end{algorithmic}
\end{algorithm}

\section{Experiments}
\label{sec:experiments}

\subsection{Data Description}
\label{subsec:data}

This study employs real trading data from CSI 300 index options in 2022 for empirical evaluation. The experimental design follows the principle of comparison under identical data splits and evaluation metrics, with results reported separately on different moneyness subsets to examine model robustness. The dataset comprises daily trading records throughout the year, covering contracts with various expiration dates, strike prices, and option types.

To ensure data quality, this paper performs systematic cleaning on the raw trading data, with the exchange settlement price adopted as the supervision label, which offers higher stability and reproducibility compared to the closing price. Regarding sample selection, contracts on the expiration date are first excluded due to their drastic price fluctuations and low trading volume. Samples with settlement prices below 0.01 CNY are then removed to avoid numerical instability from extremely low-priced options. After cleaning, the minimum settlement price in the dataset is 0.2 CNY, ensuring computational stability for relative error metrics such as MAPE.

Data splitting adopts a random shuffling strategy at the trading day level, dividing the data into training, validation, and test sets in a 7:2:1 ratio. Taking a single split as an example, the dataset contains a total of 61,086 samples, with 42,927 in the training set, 11,926 in the validation set, and 6,233 in the test set. To ensure the robustness of experimental results, this paper uses 5 different random seeds to generate 5 independent data splits, conducting complete training and evaluation procedures on each split, and ultimately reporting the mean across the 5 experiments. The rationale for adopting random splitting rather than strict chronological splitting is that this paper focuses on cross-sectional pricing problems given information available on the pricing date, rather than time-series rolling prediction problems. Additionally, market volatility characteristics vary significantly across different months in 2022, and random splitting enables the test set to cover various market conditions throughout the year, thereby providing a more comprehensive evaluation of the model's pricing capability under different volatility environments.

Based on option moneyness, this paper further divides the data into three subsets: at-the-money, in-the-money, and out-of-the-money, with specific classification criteria shown in Table~\ref{tab:moneyness}.

\begin{table}[htbp]
	\caption{Moneyness Classification Criteria and Sample Distribution}\label{tab:moneyness}
	\begin{tabular}{lccc}
		\toprule
		Moneyness & Call Options & Put Options & Sample Size \\
		\midrule
		At-the-money (ATM) & $0.97 \leq M \leq 1.03$ & $0.97 \leq M \leq 1.03$ & 10,785 \\
		In-the-money (ITM) & $M > 1.03$ & $M < 0.97$ & 23,883 \\
		Out-of-the-money (OTM) & $M < 0.97$ & $M > 1.03$ & 26,418 \\
		\bottomrule
	\end{tabular}
	\begin{tablenotes}
		\small
		\item Note: $M = X/K$ denotes the ratio of the underlying asset price $X$ to the strike price $K$.
	\end{tablenotes}
\end{table}

Furthermore, the model proposed in this paper incorporates realized volatility trajectories and market sentiment features as inputs. Realized volatility is computed based on high-frequency trading data of the CSI 300 index, and sentiment features are constructed from contemporaneous Guba post titles. Detailed methodologies are provided in Section~\ref{subsec:rv} and Section~\ref{subsec:sentiment}.

\subsection{Evaluation Metrics}
\label{subsec:metrics}

This paper employs three complementary metrics to evaluate model pricing accuracy. MAE measures the mean absolute deviation between predicted and actual values, providing an intuitive reflection of the actual magnitude of pricing errors. Root Mean Square Error (RMSE) assigns higher weights to larger errors, capturing model performance under extreme conditions. MAPE measures relative errors and is suitable for horizontal comparison across options at different price levels, being particularly sensitive to pricing quality for low-priced options. The three metrics are defined as follows:
\begin{equation}
	\mathrm{MAE} = \frac{1}{n}\sum_{i=1}^{n}\left|y_i - \hat{y}_i\right|,
\end{equation}
\begin{equation}
	\mathrm{RMSE} = \sqrt{\frac{1}{n}\sum_{i=1}^{n}\left(y_i - \hat{y}_i\right)^2},
\end{equation}
\begin{equation}
	\mathrm{MAPE} = \frac{100\%}{n}\sum_{i=1}^{n}\left|\frac{y_i - \hat{y}_i}{y_i}\right|,
\end{equation}
where $y_i$ denotes the actual option price and $\hat{y}_i$ denotes the model-predicted price.

\subsection{Experimental Setup}
\label{subsec:setup}

\subsubsection{Baseline Models}
\label{subsubsec:baseline}

This paper selects two representative baseline models for comparison. The first is the classical BSM analytical solution, adopting a risk-free rate $r=0$ and fixed volatility $\sigma$. Since the BSM model is sensitive to the volatility parameter, this paper conducts a systematic analysis of monthly pricing errors under different $\sigma$ values, with results shown in Table~\ref{tab:bsm_monthly}. Based on comprehensive annual performance, $\sigma=0.20$ achieves optimal or near-optimal pricing accuracy in most months, and is therefore adopted as the BSM baseline parameter in this paper.

\begin{table}[htbp]
	\centering
	\caption{Monthly Pricing Errors of the BSM Model under Different Volatility Parameters (MAE, Unit: CNY)}
	\label{tab:bsm_monthly}
	\begin{tabular}{lcccccc}
		\toprule
		Month & Sample Size & $\sigma$=0.10 & $\sigma$=0.15 & $\sigma$=0.20 & $\sigma$=0.25 & $\sigma$=0.30 \\
		\midrule
		2022-01 & 3808 & 50.35 & 22.91 & \textbf{14.67} & 48.49 & 86.75 \\
		2022-02 & 3515 & 37.87 & \textbf{15.87} & 20.86 & 50.70 & 85.12 \\
		2022-03 & 6144 & 57.72 & 39.70 & \textbf{29.07} & 33.90 & 54.30 \\
		2022-04 & 5262 & 60.77 & 42.57 & 25.18 & \textbf{23.66} & 43.92 \\
		2022-05 & 5353 & 54.94 & 39.72 & \textbf{27.13} & 27.06 & 43.00 \\
		2022-06 & 5697 & 58.56 & 41.17 & \textbf{28.22} & 30.43 & 54.03 \\
		2022-07 & 5638 & 59.94 & 39.56 & \textbf{22.18} & 28.32 & 57.61 \\
		2022-08 & 5805 & 43.99 & 25.35 & \textbf{12.14} & 28.63 & 57.23 \\
		2022-09 & 5089 & 40.71 & 22.25 & \textbf{7.55} & 29.38 & 58.24 \\
		2022-10 & 4176 & 41.40 & 25.94 & \textbf{8.88} & 18.96 & 42.95 \\
		2022-11 & 5500 & 44.25 & 28.98 & \textbf{10.67} & 16.34 & 40.45 \\
		2022-12 & 5099 & 39.61 & 22.98 & \textbf{11.93} & 29.43 & 55.99 \\
		\bottomrule
	\end{tabular}
\end{table}

The second baseline is the Tengbinn deep learning option pricing model \citep{CITE:Shi_2025}, with identical data splits used for fair comparison. This model utilizes only basic features such as underlying price, strike price, and time to maturity, without incorporating the volatility trajectory and sentiment features introduced in this paper.

\subsubsection{Hyperparameter Settings}
\label{subsubsec:hyperparameters}

To ensure model robustness, this paper follows complete training procedures on 5 independent data splits. Each experiment uses identical network architecture and hyperparameter configurations, differing only in the random seed for data splitting. After training, evaluation is conducted on the corresponding test set for each split, and the mean across the 5 experiments is ultimately reported.

This paper adopts a two-stage training strategy consisting of pre-training and fine-tuning. In terms of numerical implementation, the time interval $[0,\tau]$ of the BSDE is discretized into $N=32$ time steps, with $M=16$ Monte Carlo paths simulated at each step and a batch size of 128. Since the RV forecast output is a daily frequency series, this paper adopts a piecewise constant alignment approach to map it to BSDE time steps: let $D$ denote the number of remaining trading days until expiration, for the $i$-th time step $t_i = i\tau/N$, the corresponding daily index is $d(i) = \min\{\lfloor iD/N \rfloor, D-1\}$, and thus $\sigma_{t_i} = \sigma_{d(i)}^{\text{daily}}$. The pre-training stage uses mixed data from both call and put options for training, with the sentiment feature module disabled, enabling the model to focus on learning the fundamental patterns of option pricing under the FBSDE framework. This stage runs for 3,500 steps with a peak learning rate of $1 \times 10^{-4}$. The fine-tuning stage trains separately for call and put options, selectively enabling the sentiment gating mechanism based on option moneyness. This stage runs for 5,000 steps with a peak learning rate of $1 \times 10^{-5}$. The learning rate schedule adopts cosine annealing. Detailed hyperparameter configurations are shown in Table~\ref{tab:hyper}.

\begin{table}[htbp]
	\centering
	\caption{Training Hyperparameter Configuration}\label{tab:hyper}
	\begin{tabular}{lccc}
		\toprule
		Parameter & Pre-training & Fine-tuning & Description \\
		\midrule
		Batch size & 128 & 128 & Fixed \\
		Number of time steps & 32 & 32 & FBSDE discretization \\
		Number of MC paths & 16 & 16 & Monte Carlo simulation \\
		Peak learning rate & $1 \times 10^{-4}$ & $1 \times 10^{-5}$ & Cosine annealing \\
		Minimum learning rate & $1 \times 10^{-8}$ & $1 \times 10^{-8}$ & Annealing lower bound \\
		Total training steps & 3500 & 5000 & --- \\
		Warmup steps & 1400 & 500 & Linear warmup \\
		Gradient clipping & 10.0 & 20.0 & Prevent gradient explosion \\
		Weight decay & $1 \times 10^{-5}$ & $1 \times 10^{-5}$ & L2 regularization \\
		\midrule
		\multicolumn{4}{l}{\textit{Loss Function Weights}} \\
		$\lambda_1$ (price supervision) & 0.8 & 0.8 & --- \\
		$\lambda_2$ (terminal condition) & 0.1 & 0.1 & --- \\
		$\lambda_3$ (path consistency) & 0.1 & 0.1 & --- \\
		$\omega_r$ (RMSE) & 1.0 & 1.0 & --- \\
		$\omega_m$ (MAPE) & 1.0 & 1.0 & --- \\
		\midrule
		\multicolumn{4}{l}{\textit{Sentiment Gating Configuration}} \\
		Sentiment gating (ATM) & Disabled & 0.25 & --- \\
		Sentiment gating (ITM) & Disabled & Disabled & --- \\
		Sentiment gating (OTM) & Disabled & 1.2 & --- \\
		\bottomrule
	\end{tabular}
\end{table}

\subsubsection{Sentiment Gating Configuration}
\label{subsubsec:sent_gate}

For options with different moneyness levels, this paper adopts differentiated sentiment feature configuration strategies, with specific parameter settings shown in Table~\ref{tab:sentiment_gate}.

\begin{table}[htbp]
	\centering
	\caption{Sentiment Gating Configuration for Options with Different Moneyness Levels}\label{tab:sentiment_gate}
	\begin{tabular}{lcp{6.5cm}}
		\toprule
		Moneyness & Initial Value & Rationale \\
		\midrule
		ATM & 0.25 & Multi-factor sensitivity; weight learned via optimization \\
		ITM & 0 & Intrinsic value dominant; minimal sentiment exposure \\
		OTM & 1.2 & Time value dominant; elevated sentiment sensitivity \\
		\bottomrule
	\end{tabular}
\end{table}

The above design is based on fundamental financial logic of option pricing. The value of in-the-money options is primarily determined by intrinsic value, and investor decisions are more driven by fundamentals rather than market sentiment, hence the sentiment channel is disabled. Out-of-the-money options consist almost entirely of time value with high leverage, and market sentiment and speculative demand exert more significant influence on their prices, hence a higher gating initial value is assigned. At-the-money options are in a critical state, and a moderate initial value is adopted, allowing the model to autonomously adjust the contribution weight of sentiment features through gradient descent. Based on the above mechanism, the following hypothesis can be formulated: out-of-the-money and at-the-money options exhibit stronger dependence on sentiment features, while in-the-money options exhibit weaker dependence on sentiment. This hypothesis will be validated in subsequent experiments.

\subsection{Results and Analysis}
\label{subsec:results}

\subsubsection{Loss Curves}
\label{subsubsec:loss_curves}

\begin{figure}[htbp]
	\centering
	\includegraphics[width=0.90\linewidth]{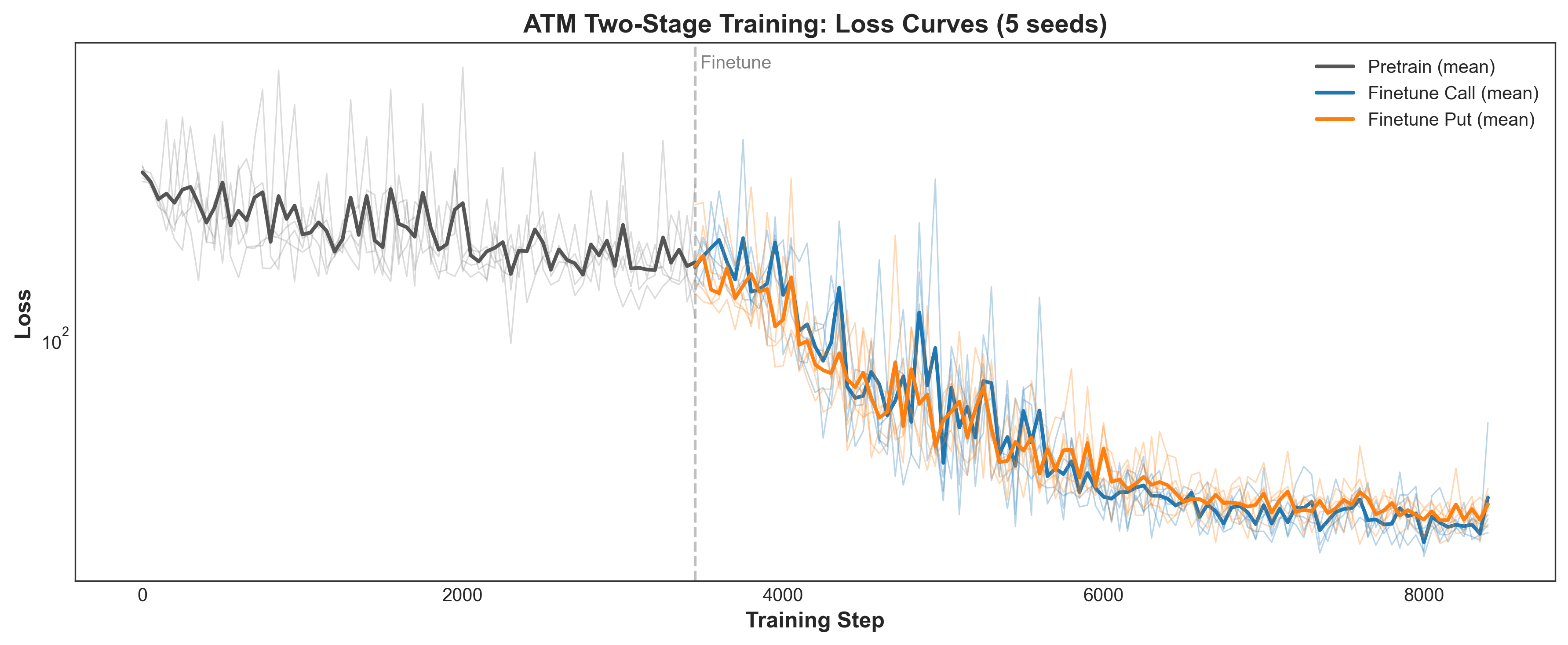}\\[4pt]
	\includegraphics[width=0.90\linewidth]{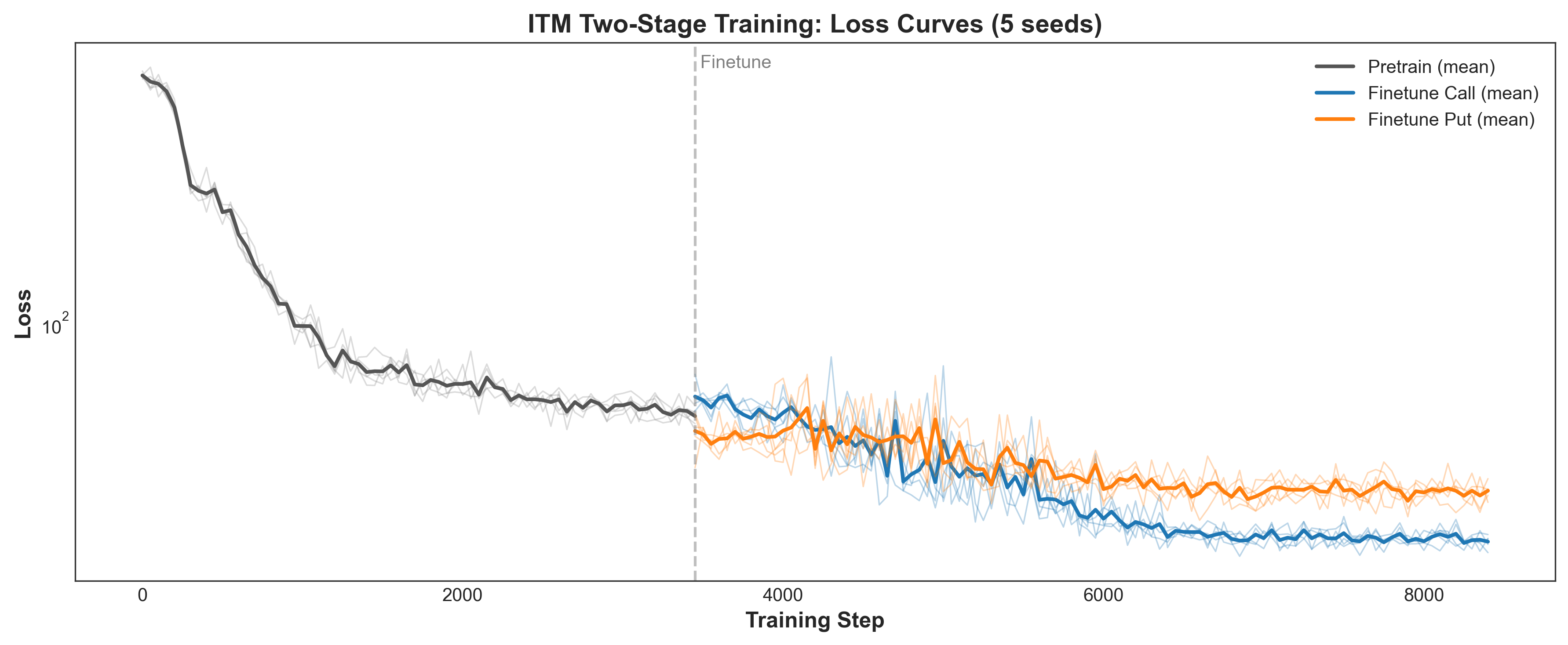}\\[4pt]
	\includegraphics[width=0.90\linewidth]{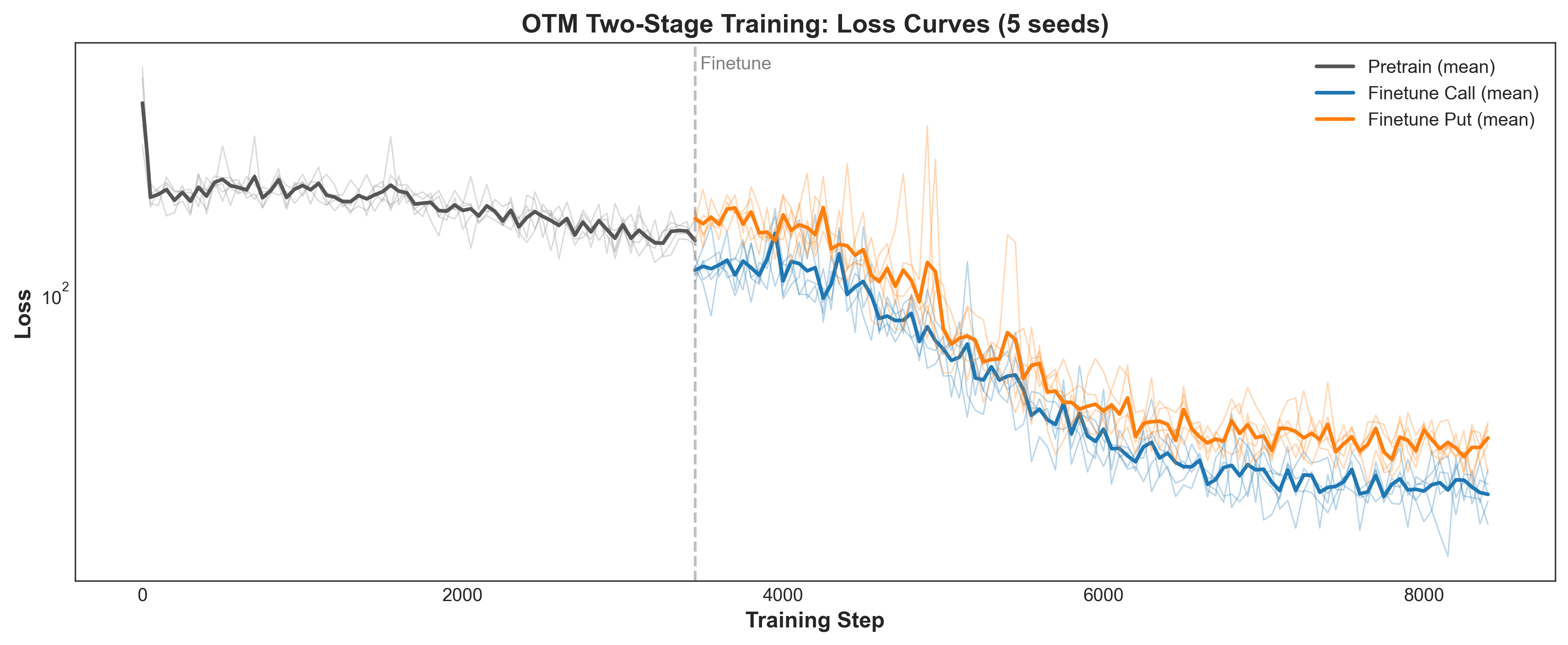}
	\caption{Loss Function Curves of Two-Stage Training}
	\label{fig:loss_curves}
\end{figure}

To validate the effectiveness of the two-stage pre-training and fine-tuning strategy, Figure~\ref{fig:loss_curves} presents the loss function convergence curves on the at-the-money (ATM), in-the-money (ITM), and out-of-the-money (OTM) subsets. In the figure, the dark curves represent the average loss across 5 independent experiments, the shaded regions indicate the loss distribution under different random seeds, and the vertical dashed line marks the transition point from the pre-training stage to the fine-tuning stage.

In terms of convergence trends, all three subsets achieved significant loss reduction during the pre-training stage. The ATM subset decreased from an initial value of 270.6 to 180.0, a reduction of approximately 33\%. The ITM subset decreased from 582.7 to 45.8, a reduction of 92\%, exhibiting the most pronounced convergence. The OTM subset decreased from 233.5 to 114.5, a reduction of approximately 51\%. Upon entering the fine-tuning stage, losses on all subsets further converged: the call and put branches of the ATM subset decreased to 30.7 and 38.5 respectively, the ITM subset to 19.6 and 27.6, and the OTM subset to 38.4 and 45.5. These results demonstrate that the pre-training stage provides favorable parameter initialization for subsequent task adaptation, and the two-stage training strategy enhances subtask fitting accuracy while maintaining convergence stability.

In terms of subset-specific performance, the ITM subset converges fastest with minimal fluctuation, consistent with the characteristic that ITM option pricing is primarily determined by intrinsic value and the pricing patterns are more readily captured by the model. The ATM subset lies in the critical region of moneyness, exhibiting stronger sample heterogeneity and relatively pronounced fluctuations during the pre-training stage, but stabilizing after fine-tuning. The OTM subset exhibits relatively stronger curve fluctuations due to the higher proportion of low-priced samples and larger error variance, yet still demonstrates a clear downward trend during the fine-tuning stage.

\subsubsection{Overall Performance}
\label{subsubsec:overall}

Table~\ref{tab:moneyness_performance} summarizes the pricing performance comparison between our method and the baseline models across overall and moneyness-specific categories.

\begin{table}[htbp]
	\centering
	\caption{Option Pricing Performance Comparison}\label{tab:moneyness_performance}
	\small
	\setlength{\tabcolsep}{4pt}
	\begin{tabular}{llrrrrrrrrr}
		\toprule
		& & \multicolumn{3}{c}{BSM} & \multicolumn{3}{c}{Tengbinn} & \multicolumn{3}{c}{Ours} \\
		\cmidrule(lr){3-5} \cmidrule(lr){6-8} \cmidrule(lr){9-11}
		Moneyness & Type & MAE & RMSE & MAPE & MAE & RMSE & MAPE & MAE & RMSE & MAPE \\
		\midrule
		Overall & All & 19.92 & 30.33 & 24.18 & 17.09 & 23.51 & 22.89 & $\mathbf{13.51}$ & $\mathbf{20.29}$ & $\mathbf{15.65}$ \\
		\midrule
		\multirow{2}{*}{ATM}
		& Call & 17.69 & 22.71 & 15.18 & 13.14 & 17.27 & 13.20 & $\mathbf{11.40}$ & $\mathbf{14.38}$ & $\mathbf{12.68}$ \\
		& Put & 26.50 & 35.90 & 16.56 & 24.41 & 30.65 & 18.37 & $\mathbf{18.68}$ & $\mathbf{24.79}$ & $\mathbf{14.90}$ \\
		\midrule
		\multirow{2}{*}{ITM}
		& Call & 23.05 & 29.78 & 5.73 & 17.27 & 21.86 & 4.52 & $\mathbf{12.44}$ & $\mathbf{16.29}$ & $\mathbf{3.32}$ \\
		& Put & 31.53 & 43.89 & 5.50 & 25.80 & 33.08 & 4.92 & $\mathbf{20.73}$ & $\mathbf{27.84}$ & $\mathbf{3.89}$ \\
		\midrule
		\multirow{2}{*}{OTM}
		& Call & 7.62 & 12.16 & 45.18 & $\mathbf{6.65}$ & $\mathbf{10.57}$ & 47.35 & 7.09 & 12.88 & $\mathbf{24.63}$ \\
		& Put & 17.29 & 26.73 & 43.99 & 15.19 & 22.39 & 48.93 & $\mathbf{12.91}$ & $\mathbf{20.83}$ & $\mathbf{31.18}$ \\
		\bottomrule
	\end{tabular}
	\begin{tablenotes}
		\footnotesize
		\item Note: MAPE is in \%, MAE and RMSE are in CNY, bold indicates the best performance. Overall metrics are computed by direct aggregation on the full test set without group-weighted averaging.
	\end{tablenotes}
\end{table}

Our method achieves the best performance across all overall metrics, reducing MAE by 32.2\% compared to BSM and by 21.0\% compared to Tengbinn. The following analysis examines the results from two dimensions: moneyness-specific performance and pricing robustness.

\subsubsection{Moneyness Analysis}
\label{subsubsec:moneyness}

The ITM segment is where our method demonstrates the most significant advantage. Our method substantially outperforms both baselines across all metrics. For call options, the MAE is reduced by 45.4\% compared to BSM and by 34.2\% compared to Tengbinn. For put options, the MAE is reduced by 33.6\% compared to BSM and by 25.4\% compared to Tengbinn. Meanwhile, the MAPE is controlled within 4\%. These results indicate that the proposed dual-network architecture and FBSDE framework can effectively capture the pricing patterns of ITM options.

In the ATM segment, our method achieves the best performance across all three metrics: MAE, RMSE, and MAPE. For call options, the MAE is reduced by 25.3\% compared to BSM and by 13.3\% compared to Tengbinn, with a MAPE of 12.68\%, outperforming BSM (15.18\%) and Tengbinn (13.20\%). For put options, the MAE is reduced by 32.8\% compared to BSM and by 20.2\% compared to Tengbinn, with a MAPE of 14.90\%, outperforming BSM (16.56\%) and Tengbinn (18.37\%). ATM options lie in the critical region of moneyness, where pricing is simultaneously influenced by both intrinsic value and time value. By integrating volatility trajectory and sentiment features, our method better captures the pricing patterns in this segment.

The OTM segment presents higher pricing challenges. Due to lower option prices, absolute errors tend to be small, making relative errors more discriminative for pricing quality assessment. In terms of MAE, our method achieves the best performance for OTM put options, reducing MAE by 15.0\% compared to Tengbinn. The key difference lies in the MAPE metric: our method controls MAPE within the range of 25\% to 31\%, whereas Tengbinn yields approximately 47\% to 49\% and BSM approximately 44\% to 45\%. This demonstrates that our method possesses significant advantages in relative error control, with the high sentiment gating initial value strategy described in Section~\ref{subsubsec:sent_gate} playing a crucial role.

\subsubsection{Robustness Analysis}
\label{subsubsec:robustness}

Due to their lower prices, OTM options serve as a critical test scenario for pricing robustness. This section analyzes the differences between our method and the deep learning baseline (Tengbinn) from the perspective of extreme error control. Table~\ref{tab:extreme_error} presents the statistics on the proportion of extreme error samples in the OTM segment.

\begin{table}[htbp]
	\centering
	\caption{Proportion of Extreme Error Samples in the OTM Segment}\label{tab:extreme_error}
	\begin{tabular}{lccc}
		\toprule
		Model & MAPE$>$50\% & MAPE$>$100\% & MAPE$>$200\% \\
		\midrule
		Tengbinn & 26.82\% & 11.51\% & 4.24\% \\
		Ours & \textbf{8.74\%} & \textbf{0.77\%} & \textbf{0.19\%} \\
		\bottomrule
	\end{tabular}
\end{table}

Our method demonstrates significant advantages in extreme error control. Tengbinn has 11.51\% of OTM samples with MAPE exceeding 100\%, whereas our method has only 0.77\%, a reduction of approximately 93\%.

OTM low-priced options are extremely sensitive to pricing deviations, where minor absolute errors can lead to substantial relative errors. Through the physical constraints of the FBSDE framework and the sentiment gating strategy, our method effectively suppresses systematic pricing biases, thereby substantially reducing the occurrence of extreme error samples.

\subsubsection{Visualization}
\label{subsubsec:visual}

\begin{figure}[htbp]
	\centering
	\includegraphics[width=0.98\linewidth]{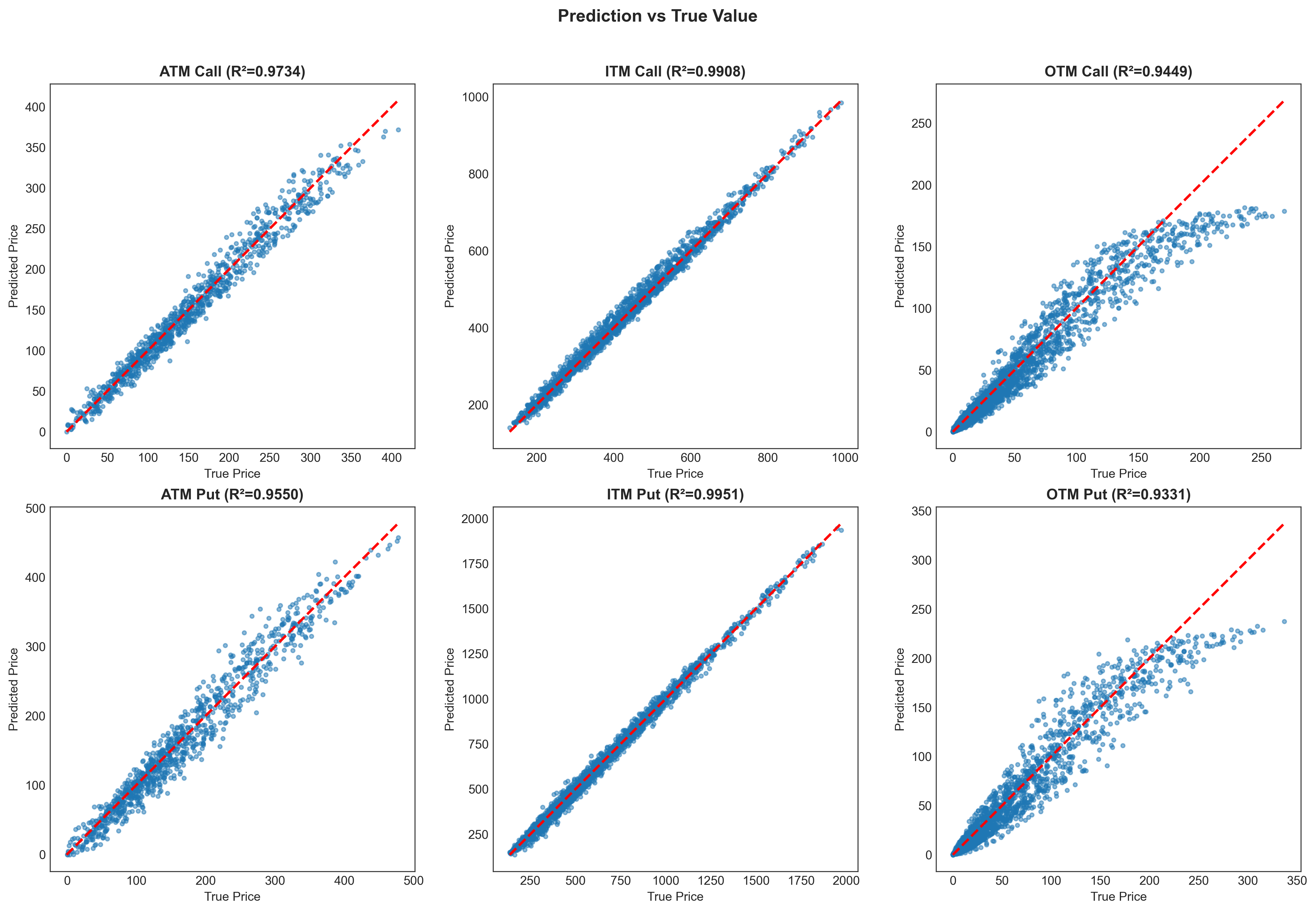}
	\caption{Scatter Plots of Predicted versus Actual Values across Different Moneyness Levels}
	\label{fig:pred_scatter}
\end{figure}

Figure~\ref{fig:pred_scatter} visually illustrates the fitting consistency of the model across different moneyness levels and option types from the perspective of predicted versus actual values. Overall, the point cloud of the ITM subset lies closest to the diagonal line $y=x$, corresponding to the highest goodness of fit, with $R^2=0.9908$ for call options and $R^2=0.9951$ for put options. The ATM subset follows, with $R^2=0.9734$ for call options and $R^2=0.9550$ for put options. The OTM subset exhibits the largest dispersion, with $R^2=0.9449$ for call options and $R^2=0.9331$ for put options.

In terms of bias patterns, the OTM subset exhibits a certain degree of underestimation in the region of larger actual prices, with some points lying below the diagonal line. In contrast, the point clouds of the ITM and ATM subsets are evenly distributed on both sides of the diagonal, indicating that the model possesses favorable fitting stability in these two segments. In summary, our method achieves optimal performance on overall metrics and most moneyness-specific metrics, while maintaining good fitting consistency across all moneyness levels.

\section{Interpretability and Economic Implications}
\label{sec:xai}

\subsection{Architecture Advantage Analysis}
\label{subsec:arch}

The performance improvement of our method relative to baseline models may stem from two sources: the nonlinear fitting capability of the deep learning architecture itself, and the ability to utilize additional market information. This paper refers to the former as architectural advantage and the latter as information advantage. To isolate the architectural advantage, this paper designs an ablation experiment by setting the volatility trajectory to a constant value of 0.2 and the sentiment gating parameter to 0, ensuring that the model uses the same information inputs as BSM and Tengbinn.

Since BSM is an analytical solution and Tengbinn is a deep BSDE method, this paper defines two types of architectural advantage to distinguish different comparison contexts:
\begin{align}
	\Delta_{\text{arch}}^{\text{BSM}} &= \text{MAE}_B - \text{MAE}_C, \\
	\Delta_{\text{arch}}^{\text{TB}} &= \text{MAE}_{TB} - \text{MAE}_C,
\end{align}
where $\Delta_{\text{arch}}^{\text{BSM}}$ measures the architectural advantage of our method relative to BSM, $\Delta_{\text{arch}}^{\text{TB}}$ measures the architectural advantage relative to Tengbinn, and positive values indicate that our method outperforms the corresponding baseline under identical information conditions. This paper employs both MAE and MAPE metrics to measure improvements in absolute and relative errors, respectively.

\begin{table}[htbp]
	\centering
	\caption{Architectural Advantage Analysis}\label{tab:arch_advantage}
	\begin{tabular}{lcccccccc}
		\toprule
		& \multicolumn{4}{c}{MAE (CNY)} & \multicolumn{4}{c}{MAPE (pp)} \\
		\cmidrule(lr){2-5} \cmidrule(lr){6-9}
		Option Type & $E_A$ & $E_C$ & $\Delta_{\text{arch}}^{\text{BSM}}$ & $\Delta_{\text{arch}}^{\text{TB}}$ & $E_A$ & $E_C$ & $\Delta_{\text{arch}}^{\text{BSM}}$ & $\Delta_{\text{arch}}^{\text{TB}}$ \\
		\midrule
		ATM Call & 11.40 & 11.53 & 6.13 & 1.51 & 12.7 & 13.1 & 1.8 & $-$0.2 \\
		ATM Put & 18.69 & 19.41 & 7.15 & 5.03 & 14.9 & 16.0 & 0.3 & 2.1 \\
		ITM Call & 12.42 & 12.42 & 10.68 & 4.32 & 3.3 & 3.3 & 2.4 & 1.0 \\
		ITM Put & 20.74 & 20.90 & 10.78 & 4.96 & 3.9 & 3.9 & 1.6 & 1.0 \\
		OTM Call & 7.17 & 7.22 & 0.48 & $-$0.58 & 24.6 & 25.2 & 19.3 & 21.0 \\
		OTM Put & 13.01 & 13.13 & 4.07 & 0.48 & 30.9 & 31.5 & 11.9 & 19.1 \\
		\bottomrule
	\end{tabular}
	\begin{tablenotes}
		\footnotesize
		\item Note: pp = percentage points. $E_A$ denotes the error of our method, $E_C$ denotes the ablation experiment error. $\Delta_{\text{arch}}^{\text{BSM}}=E_B-E_C$, $\Delta_{\text{arch}}^{\text{TB}}=E_{TB}-E_C$. The errors of BSM and Tengbinn under identical information conditions can be directly reconstructed as $E_B = E_C + \Delta_{\text{arch}}^{\text{BSM}}$ and $E_{TB} = E_C + \Delta_{\text{arch}}^{\text{TB}}$.
	\end{tablenotes}
\end{table}

Table~\ref{tab:arch_advantage} presents the architectural advantage analysis based on both MAE and MAPE metrics. Under the MAE metric, ITM options exhibit the largest architectural advantage, with $\Delta_{\text{arch}}^{\text{BSM}}$ approximately 10 CNY and $\Delta_{\text{arch}}^{\text{TB}}$ approximately 4 to 5 CNY, followed by ATM options, with OTM options showing the smallest advantage. ITM options are primarily determined by intrinsic value with relatively clear pricing patterns, and our method achieves significant improvements over both baselines by more precisely fitting these pricing patterns through the dual-network architecture and physical constraints of the FBSDE framework. Under the MAPE metric, OTM options demonstrate the largest architectural advantage, with $\Delta_{\text{arch}}^{\text{BSM}}$ reaching 19.3 and 11.9 percentage points, and $\Delta_{\text{arch}}^{\text{TB}}$ reaching 21.0 and 19.1 percentage points. OTM options have lower prices and consist almost entirely of time value, making them highly sensitive to volatility and market expectations. Our method captures nonlinear pricing patterns through the adaptive activation function network while employing a hybrid loss function combining RMSE and MAPE that emphasizes relative error optimization, thereby achieving significant improvements in the MAPE metric for OTM options. Overall, our method demonstrates positive architectural advantage over BSM across all option types and maintains advantages over Tengbinn for most option types, indicating that the proposed network architecture possesses superior pricing capability.

\subsection{Informational Advantage Analysis}
\label{subsec:info}

\subsubsection{Integrated Gradients Attribution}
\label{subsubsec:ig_attribution}

The analysis in the previous section demonstrates that our method possesses significant architectural advantage, while information advantage is equally an important source of performance improvement. This section further analyzes the specific composition of information advantage, namely the respective contributions of volatility trajectory and market sentiment. This paper employs the Integrated Gradients method \citep{CITE:Sundararajan_2017_IG} to approximate the path integral of gradients of the network output with respect to inputs. The baseline input $x'$ is constructed as follows: the volatility trajectory input is replaced with a constant value of 0.2, and the sentiment channel is disabled (i.e., the sentiment gating coefficient is set to 0), while other inputs remain unchanged. For the $i$-th feature, the attribution value is defined as:
\begin{equation}
	\text{IG}_i(x) = (x_i - x'_i) \times \int_{\alpha=0}^{1} \frac{\partial f(x' + \alpha(x - x'))}{\partial x_i} d\alpha,
\end{equation}
where $f$ denotes the model output function. The integral is approximated discretely, with linear interpolation along the path from $x'$ to the actual input $x$, using $m=10$ discrete steps to compute the gradient accumulation. To facilitate comparison across samples and features, this paper converts the attribution results into percentage form: the absolute values of attributions for each feature are summed and normalized by the total absolute attribution, yielding the relative contribution proportion of each component.

Table~\ref{tab:ig_attribution} presents the Integrated Gradients attribution results for different option types. Since the sentiment channel is disabled during training for ITM options, their sentiment contribution is zero, and thus only the attribution results for ATM and OTM options are reported in the table. The results show that ATM call options have the highest sentiment contribution at 60.5\%, while put options generally exhibit volatility trajectory dominance, with OTM put options showing a volatility trajectory attribution of 62.2\%.

\begin{table}[htbp]
	\centering
	\caption{Integrated Gradients Attribution Results}\label{tab:ig_attribution}
	\begin{tabular}{lcc}
		\toprule
		Option Type & Volatility trajectory Contribution & Sentiment Contribution \\
		\midrule
		ATM Call & 39.5\% & 60.5\% \\
		ATM Put & 50.2\% & 49.8\% \\
		OTM Call & 55.4\% & 44.6\% \\
		OTM Put & 62.2\% & 37.8\% \\
		\bottomrule
	\end{tabular}
\end{table}

\subsubsection{Shapley Attribution}
\label{subsubsec:shapley_attribution}

The Integrated Gradients method analyzes the sensitivity of model output to input features, while the Shapley value method analyzes the contribution of each feature from the perspective of error improvement. This paper employs a Shapley value decomposition method based on model ablation, which only replaces or masks corresponding inputs during the inference phase without modifying model parameters, constructing four input configurations: baseline configuration (volatility trajectory set to constant 0.2, sentiment gating set to 0), volatility trajectory only, sentiment only, and full configuration. All four configurations are evaluated on the same trained model without retraining, yielding corresponding MAE values denoted as $E_{\text{none}}$, $E_{\text{rv}}$, $E_{\text{sent}}$, and $E_{\text{full}}$. Based on cooperative game theory \citep{CITE:Shapley_1953}, the contribution of each information source can be computed via Shapley values:
\begin{align}
	\phi_{\text{RV}} &= \frac{1}{2}\left[(E_{\text{none}} - E_{\text{rv}}) + (E_{\text{sent}} - E_{\text{full}})\right], \\
	\phi_{\text{Sent}} &= \frac{1}{2}\left[(E_{\text{none}} - E_{\text{sent}}) + (E_{\text{rv}} - E_{\text{full}})\right],
\end{align}
where the $\phi$ values are defined as the MAE reduction relative to the corresponding ablation configuration, measured in CNY. Through combinatorial computation of error changes under different input configurations, the overall error improvement can be additively allocated among information sources.

\begin{figure}[htbp]
	\centering
	\includegraphics[width=0.9\textwidth]{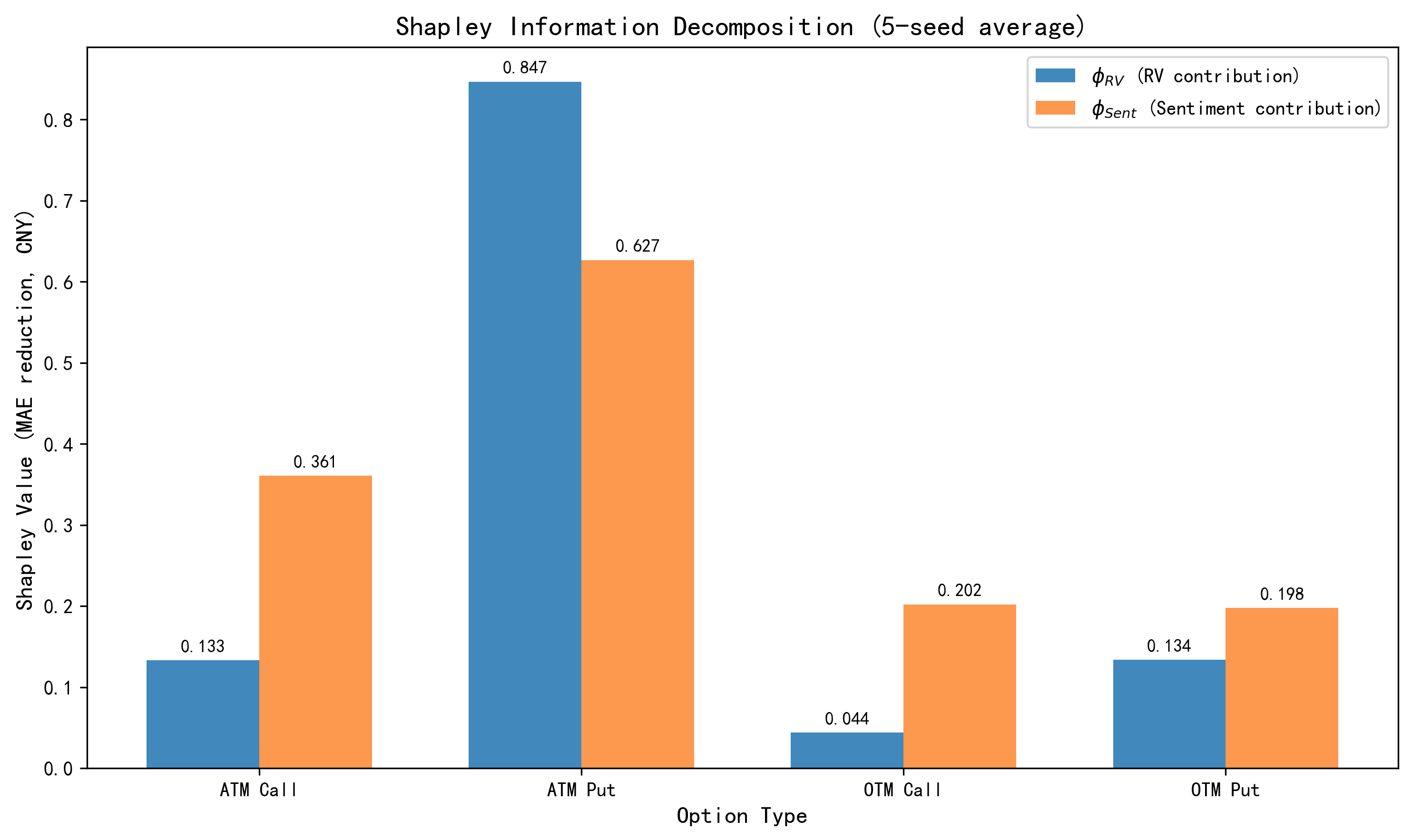}
	\caption{Shapley Value Decomposition Results}
	\label{fig:shapley_decomposition}
\end{figure}

Figure~\ref{fig:shapley_decomposition} presents the Shapley value decomposition results in bar chart form, visually illustrating the differences in information sources between call and put options. Since the sentiment channel is disabled during training for ITM options, only the decomposition results for ATM and OTM options are shown. It can be observed that the orange bars (sentiment contribution) for call options are significantly higher than the blue bars (volatility trajectory contribution), while the two contributions for put options are more balanced.

Table~\ref{tab:shapley_decomposition} provides detailed quantitative analysis results. In terms of total information gain, all option types exhibit positive total information gain, indicating that the introduction of volatility trajectory and sentiment information contributes positively to the pricing of all option categories. Among them, ATM put options show the largest total information gain of 1.474 CNY, suggesting that this option type benefits most from the additional information.

\begin{table}[htbp]
	\centering
	\caption{Shapley Value Decomposition Results}\label{tab:shapley_decomposition}
	\begin{tabular}{lccccc}
		\toprule
		Option Type & $\phi_{\text{RV}}$ & $\phi_{\text{Sent}}$ & RV Share & Sent Share & Total Gain \\
		\midrule
		ATM Call & 0.133 & 0.361 & 9.3\% & 90.7\% & 0.494 \\
		ATM Put & 0.847 & 0.627 & 60.9\% & 39.1\% & 1.474 \\
		OTM Call & 0.044 & 0.202 & 29.6\% & 70.4\% & 0.246 \\
		OTM Put & 0.134 & 0.198 & 44.5\% & 55.5\% & 0.332 \\
		\bottomrule
	\end{tabular}
	\begin{tablenotes}
		\footnotesize
		\item Note: $\phi$ values represent MAE reduction (CNY). RV = volatility trajectory, Sent = sentiment. Total Gain = $\phi_{\text{RV}} + \phi_{\text{Sent}}$.
	\end{tablenotes}
\end{table}

In terms of contribution share, the error improvement for call options is primarily driven by market sentiment, with sentiment contribution reaching 90.7\% for ATM calls and 70.4\% for OTM calls. In contrast, the contributions of volatility trajectory and sentiment are more balanced for put options, with volatility trajectory contributing 60.9\% for ATM puts and sentiment slightly higher at 55.5\% for OTM puts.

The above asymmetry pattern between calls and puts is consistent with the conclusions from Integrated Gradients attribution. The two methods mutually validate each other from different perspectives, enhancing the reliability of the conclusions. This asymmetry carries clear economic implications. From the perspective of risk exposure, call options reflect bullish expectations and are most sensitive to upward movements in the underlying price. Consequently, their pricing is more susceptible to factors such as investor sentiment and changes in risk appetite. When market sentiment is elevated, investors assign higher subjective probabilities to underlying price appreciation, leading to increased demand for call options and subsequently higher prices. From the perspective of market demand, put options are primarily used for tail risk hedging. Institutional investors typically purchase put options for risk management purposes, and their pricing is more dominated by the level of market uncertainty, thus exhibiting stronger dependence on volatility trajectory information. Furthermore, the total information gain for ATM put options is substantially higher than for other types, which may reflect the robust hedging demand in this segment, where both volatility trajectory and sentiment information provide effective supplementary inputs for pricing.

\subsection{Managerial Implications}
The interpretability analysis yields several practical insights for derivatives practitioners and risk managers.

First, the asymmetric information dependence between calls and puts suggests differentiated pricing strategies: call option traders should prioritize real-time sentiment monitoring, while put option pricing should emphasize volatility trajectory forecasting. This finding can inform the design of trading algorithms and hedging strategies tailored to option types.

Second, the substantial reduction in extreme pricing errors (MAPE$>$100\% reduced from 11.51\% to 0.77\% for OTM options) has direct implications for margin calculations and counterparty risk management, particularly for low-priced options that are prone to mispricing under traditional models.

Third, the sentiment gating configuration (Table~\ref{tab:sentiment_gate}) provides actionable guidance for model deployment: practitioners may assign higher sentiment weights to OTM options and lower weights to ITM options, aligning model parameters with their respective pricing mechanisms. This adaptive weighting scheme can be integrated into existing pricing systems with minimal modification.

Finally, the deep FBSDE framework offers a transparent and interpretable alternative to black-box machine learning models, facilitating regulatory compliance and model validation in financial institutions.

\section{Conclusion}
\label{sec:conclusion}
This paper addresses the issues of constant volatility assumption, absence of behavioral factors, and insufficient model interpretability in option pricing by proposing an option pricing method based on the deep FBSDE framework. The method integrates volatility trajectory and market sentiment features, achieves joint learning of the value function and generator through a dual-network architecture, enhances nonlinear expressive power via adaptive activation function networks, and realizes adaptive information fusion through differentiated sentiment gating strategies. Experiments on the CSI 300 index options dataset demonstrate that our method reduces MAE by 32.2\% and MAPE by 35.3\% compared to the BSM baseline, achieving optimal performance on overall metrics and most moneyness-specific metrics. Meanwhile, interpretability analysis reveals two important findings. First, the proposed network architecture yields significant pricing accuracy improvements over baseline models under identical information conditions. Second, call and put options exhibit asymmetry in their information dependence patterns. The interpretability analysis framework proposed in this paper, including architectural advantage decomposition, Integrated Gradients attribution, and Shapley value decomposition, can be extended to the interpretation and diagnosis of other deep learning pricing models.

\section*{Data availability statement}
Data availability is restricted due to licensing agreements with data providers. Code is available from the corresponding author upon reasonable request.

\section*{Funding}
This work was supported by the National Key R\&D Program of China (Grant No. 2023YFA1008903), the Major Fundamental Research Project of Shandong Province of China (Grant No. ZR2023ZD33), Taishan Scholar Project of Shandong Province of China (Grant tstp20240803), the Natural Science Foundation of Shandong Province (Grant No. ZR2023LLZ012), the Key Project of Shandong Provincial Key R\&D Program (Soft Science) (Grant No. 2024RZB0204).

\section*{Conflict of interest}
The authors declare that they have no conflict of interest.

\bibliographystyle{plainnat}
\bibliography{ref}

@article{CITE:Agostinelli_2014,
  author  = {Agostinelli, Forest and Hoffman, Matthew and Sadowski, Peter and Baldi, Pierre},
  title   = {Learning Activation Functions to Improve Deep Neural Networks},
  year    = {2014},
  journal = {arXiv preprint arXiv:1412.6830}
}

@article{CITE:Andersen_2003,
  author  = {Andersen, Torben G. and Bollerslev, Tim and Diebold, Francis X. and Labys, Paul},
  title   = {Modeling and Forecasting Realized Volatility},
  journal = {Econometrica},
  volume  = {71},
  number  = {2},
  pages   = {579--625},
  year    = {2003}
}

@article{CITE:Andersen_Bollerslev_1998,
  author  = {Andersen, Torben G. and Bollerslev, Tim},
  title   = {Answering the Skeptics: Yes, Standard Volatility Models Do Provide Accurate Forecasts},
  journal = {International Economic Review},
  volume  = {39},
  number  = {4},
  pages   = {885--905},
  year    = {1998}
}

@article{CITE:Antweiler_Frank_2004,
  author  = {Antweiler, Werner and Frank, Murray Z.},
  title   = {Is All That Talk Just Noise? The Information Content of Internet Stock Message Boards},
  journal = {Journal of Finance},
  volume  = {59},
  number  = {3},
  pages   = {1259--1294},
  year    = {2004}
}

@article{CITE:Baker_Wurgler_2006,
  author  = {Baker, Malcolm and Wurgler, Jeffrey},
  title   = {Investor Sentiment and the Cross-Section of Stock Returns},
  journal = {Journal of Finance},
  volume  = {61},
  number  = {4},
  pages   = {1645--1680},
  year    = {2006}
}

@article{CITE:Barles_Soner_1998,
  author  = {Barles, Guy and Soner, Halil Mete},
  title   = {Option Pricing with Transaction Costs and a Nonlinear Black-Scholes Equation},
  journal = {Finance and Stochastics},
  volume  = {2},
  number  = {4},
  pages   = {369--397},
  year    = {1998}
}

@article{CITE:Barndorff_Nielsen_2002,
  author  = {Barndorff-Nielsen, Ole E. and Shephard, Neil},
  title   = {Econometric Analysis of Realized Volatility and Its Use in Estimating Stochastic Volatility Models},
  journal = {Journal of the Royal Statistical Society: Series B},
  volume  = {64},
  number  = {2},
  pages   = {253--280},
  year    = {2002}
}

@inproceedings{CITE:Bengio_2009_Curriculum,
  author    = {Bengio, Yoshua and Louradour, Jérôme and Collobert, Ronan and Weston, Jason},
  title     = {Curriculum Learning},
  booktitle = {Proceedings of the 26th International Conference on Machine Learning (ICML)},
  pages     = {41--48},
  year      = {2009}
}

@article{CITE:Black_Scholes_1973,
  author  = {Black, Fischer and Scholes, Myron},
  title   = {The Pricing of Options and Corporate Liabilities},
  journal = {Journal of Political Economy},
  volume  = {81},
  number  = {3},
  pages   = {637--654},
  year    = {1973}
}

@article{CITE:Cetin_2004,
  author  = {{\c{C}}etin, Umut and Jarrow, Robert A. and Protter, Philip},
  title   = {Liquidity Risk and Arbitrage Pricing Theory},
  journal = {Finance and Stochastics},
  volume  = {8},
  number  = {3},
  pages   = {311--341},
  year    = {2004}
}

@article{CITE:Cont_2001,
  author  = {Cont, Rama},
  title   = {Empirical Properties of Asset Returns: Stylized Facts and Statistical Issues},
  journal = {Quantitative Finance},
  volume  = {1},
  number  = {2},
  pages   = {223--236},
  year    = {2001}
}

@article{CITE:Derman_Kani_1994,
  author  = {Derman, Emanuel and Kani, Iraj},
  title   = {Riding on a Smile},
  journal = {Risk},
  volume  = {7},
  number  = {2},
  pages   = {32--39},
  year    = {1994}
}

@inproceedings{CITE:Devlin_2019_BERT,
  author    = {Devlin, Jacob and Chang, Ming-Wei and Lee, Kenton and Toutanova, Kristina},
  title     = {BERT: Pre-training of Deep Bidirectional Transformers for Language Understanding},
  booktitle = {Proceedings of NAACL-HLT},
  year      = {2019}
}

@article{CITE:Dumas_1998,
  author  = {Dumas, Bernard and Fleming, Jeff and Whaley, Robert E.},
  title   = {Implied Volatility Functions: Empirical Tests},
  journal = {Journal of Finance},
  volume  = {53},
  number  = {6},
  pages   = {2059--2106},
  year    = {1998}
}

@article{CITE:Dupire_1994,
  author  = {Dupire, Bruno},
  title   = {Pricing with a Smile},
  journal = {Risk},
  volume  = {7},
  number  = {1},
  pages   = {18--20},
  year    = {1994}
}

@article{CITE:Feynman_1948,
  author  = {Feynman, Richard P.},
  title   = {Space-Time Approach to Non-Relativistic Quantum Mechanics},
  journal = {Reviews of Modern Physics},
  volume  = {20},
  number  = {2},
  pages   = {367--387},
  year    = {1948}
}

@book{CITE:Gatheral_2006,
  author    = {Gatheral, Jim},
  title     = {The Volatility Surface: A Practitioner's Guide},
  publisher = {Wiley},
  year      = {2006},
  isbn      = {978-0-471-79251-2}
}

@inproceedings{CITE:Glorot_Bengio_2010_Xavier,
  author    = {Glorot, Xavier and Bengio, Yoshua},
  title     = {Understanding the Difficulty of Training Deep Feedforward Neural Networks},
  booktitle = {Proceedings of AISTATS},
  year      = {2010}
}

@article{CITE:Han_2018,
  author  = {Han, Jiequn and Jentzen, Arnulf and E, Weinan},
  title   = {Solving High-Dimensional Partial Differential Equations Using Deep Learning},
  journal = {Proceedings of the National Academy of Sciences (PNAS)},
  volume  = {115},
  number  = {34},
  pages   = {8505--8510},
  year    = {2018}
}

@article{CITE:Han_Jentzen_2017,
  author  = {E, Weinan and Han, Jiequn and Jentzen, Arnulf},
  title   = {Deep learning-based numerical methods for high-dimensional parabolic partial differential equations and backward stochastic differential equations},
  year    = {2017},
  journal = {arXiv preprint arXiv:1706.04702}
}

@article{CITE:Hendrycks_Gimpel_2016_GELU,
  author  = {Hendrycks, Dan and Gimpel, Kevin},
  title   = {Gaussian Error Linear Units (GELUs)},
  year    = {2016},
  journal = {arXiv preprint arXiv:1606.08415}
}

@article{CITE:Heston_1993,
  author  = {Heston, Steven L.},
  title   = {A Closed-Form Solution for Options with Stochastic Volatility with Applications to Bond and Currency Options},
  journal = {The Review of Financial Studies},
  volume  = {6},
  number  = {2},
  pages   = {327--343},
  year    = {1993}
}

@article{CITE:Hull_White_1987,
  author  = {Hull, John and White, Alan},
  title   = {The Pricing of Options on Assets with Stochastic Volatilities},
  journal = {The Journal of Finance},
  volume  = {42},
  number  = {2},
  pages   = {281--300},
  year    = {1987}
}

@article{CITE:Hure_2020,
  author  = {Hur{\'e}, C{\^o}me and Pham, Huy{\^e}n and Warin, Xavier},
  title   = {Deep Backward Schemes for High-Dimensional Nonlinear PDEs},
  journal = {Mathematics of Computation},
  volume  = {89},
  number  = {324},
  pages   = {1547--1579},
  year    = {2020}
}

@article{CITE:Kac_1949,
  author  = {Kac, Mark},
  title   = {On Distributions of Certain Wiener Functionals},
  journal = {Transactions of the American Mathematical Society},
  volume  = {65},
  number  = {1},
  pages   = {1--13},
  year    = {1949}
}

@article{CITE:Karoui_1997,
  author  = {El Karoui, Nicole and Peng, Shige and Quenez, Marie-Claire},
  title   = {Backward Stochastic Differential Equations in Finance},
  journal = {Mathematical Finance},
  volume  = {7},
  number  = {1},
  pages   = {1--71},
  year    = {1997}
}

@book{CITE:Kloeden_Platen_1992,
  author    = {Kloeden, Peter E. and Platen, Eckhard},
  title     = {Numerical Solution of Stochastic Differential Equations},
  publisher = {Springer},
  year      = {1992},
  isbn      = {978-3-540-54062-5}
}

@article{CITE:LeCun_1998_CNN,
  author  = {LeCun, Yann and Bottou, L{\'e}on and Bengio, Yoshua and Haffner, Patrick},
  title   = {Gradient-Based Learning Applied to Document Recognition},
  journal = {Proceedings of the IEEE},
  volume  = {86},
  number  = {11},
  pages   = {2278--2324},
  year    = {1998}
}

@article{CITE:Leland_1985,
  author  = {Leland, Hayne E.},
  title   = {Option Pricing and Replication with Transactions Costs},
  journal = {Journal of Finance},
  volume  = {40},
  number  = {5},
  pages   = {1283--1301},
  year    = {1985}
}

@article{CITE:Merton_1973,
  author  = {Merton, Robert C.},
  title   = {Theory of Rational Option Pricing},
  journal = {The Bell Journal of Economics and Management Science},
  volume  = {4},
  number  = {1},
  pages   = {141--183},
  year    = {1973}
}

@article{CITE:Pardoux_Peng_1990,
  author  = {Pardoux, Etienne and Peng, Shige},
  title   = {Adapted Solution of a Backward Stochastic Differential Equation},
  journal = {Systems \& Control Letters},
  volume  = {14},
  number  = {1},
  pages   = {55--61},
  year    = {1990}
}

@incollection{CITE:Peng_1997,
  author    = {Peng, Shige},
  title     = {Backward SDE and Related $g$-Expectation},
  booktitle = {Backward Stochastic Differential Equations},
  series    = {Pitman Research Notes in Mathematics Series},
  volume    = {364},
  year      = {1997}
}

@incollection{CITE:Peng_g_expectation,
  author    = {Peng, Shige},
  title     = {Nonlinear Expectations, Nonlinear Evaluations and Risk Measures},
  booktitle = {Stochastic Methods in Finance},
  series    = {Lecture Notes in Mathematics},
  volume    = {1856},
  pages     = {165--253},
  publisher = {Springer},
  year      = {2004}
}

@article{CITE:Ramachandran_2017_Swish,
  author  = {Ramachandran, Prajit and Zoph, Barret and Le, Quoc V.},
  title   = {Searching for Activation Functions},
  year    = {2017},
  journal = {arXiv preprint arXiv:1710.05941}
}

@article{CITE:Rubinstein_1994,
  author  = {Rubinstein, Mark},
  title   = {Implied Binomial Trees},
  journal = {Journal of Finance},
  volume  = {49},
  number  = {3},
  pages   = {771--818},
  year    = {1994}
}

@incollection{CITE:Shapley_1953,
  author    = {Shapley, Lloyd S.},
  title     = {A Value for $n$-Person Games},
  booktitle = {Contributions to the Theory of Games},
  volume    = {II},
  publisher = {Princeton University Press},
  year      = {1953}
}

@article{CITE:Shi_2025,
  author  = {Shi, Yufeng and Teng, Bin and Wang, Sicong},
  title   = {Option Pricing Mechanisms Driven by Backward Stochastic Differential Equations},
  journal = {Financial Innovation},
  volume  = {11},
  number  = {1},
  pages   = {90},
  year    = {2025},
  doi     = {10.1186/s40854-024-00714-3}
}

@article{CITE:Song_2025,
  author  = {Song, Yuping and Zhang, Yilun and Huang, Jiefei and Yang, Aijun},
  title   = {Volatility and Value-at-Risk Forecasting Using BERT and Transformer Models Incorporating Investors' Textual Sentiments},
  journal = {Finance Research Letters},
  pages   = {108210},
  year    = {2025}
}

@inproceedings{CITE:Sundararajan_2017_IG,
  author    = {Sundararajan, Mukund and Taly, Ankur and Yan, Qiqi},
  title     = {Axiomatic Attribution for Deep Networks},
  booktitle = {Proceedings of the 34th International Conference on Machine Learning (ICML)},
  year      = {2017}
}

@article{CITE:Tetlock_2007,
  author  = {Tetlock, Paul C.},
  title   = {Giving Content to Investor Sentiment: The Role of Media in the Stock Market},
  journal = {Journal of Finance},
  volume  = {62},
  number  = {3},
  pages   = {1139--1168},
  year    = {2007}
}

@inproceedings{CITE:Vaswani_2017_Attention,
  author    = {Vaswani, Ashish and Shazeer, Noam and Parmar, Niki and Uszkoreit, Jakob and Jones, Llion and Gomez, Aidan N. and Kaiser, {\L}ukasz and Polosukhin, Illia},
  title     = {Attention Is All You Need},
  booktitle = {Advances in Neural Information Processing Systems (NeurIPS)},
  year      = {2017}
}

@inproceedings{CITE:Wang_TimeXer_2024,
  author    = {Wang, Yuxuan and Wu, Haixu and Dong, Jiaxiang and Liu, Yong and Qiu, Yunzhong and Zhang, Haoran and Wang, Jianmin and Long, Mingsheng},
  title     = {TimeXer: Empowering Transformers for Time Series Forecasting with Exogenous Variables},
  booktitle = {Advances in Neural Information Processing Systems (NeurIPS)},
  year      = {2024}
}

@unpublished{CITE:Zhang_2025,
  author = {Zhang, Yilun and Liu, Shiji and Shi, Yufeng},
  title  = {Multimodal Stock Volatility Forecasting across Multiple Timeframes via {TimeXer}},
  note   = {Working paper},
  year   = {2025}
}

\end{document}